\documentclass[amsmath,amssymb,twocolumn,nofootinbib,superscriptaddress]{revtex4-1}
\usepackage{color}
\usepackage[dvipsnames]{xcolor}
\usepackage{graphicx,epsfig,hyperref,latexsym}
\usepackage{amsmath}
\usepackage{braket}
\usepackage{tensor}
\usepackage[normalem]{ulem}
\usepackage{mathtools}
\usepackage{mathrsfs}

\newcommand{\eq}[1]{\begin{equation}\begin{aligned}#1\end{aligned}\end{equation}}
\newcommand{\eref}[1]{(\ref{eq:#1})}
\newcommand{\del}[2]{\frac{\partial#1}{\partial#2}}
\newcommand{\D}{\mathrm{d}}
\newcommand{\Del}[2]{\frac{\D#1}{\D#2}}

\newcommand{\I}{\mathrm{i}}
\newcommand{\A}{\mathcal{A}}

\newcommand{\Def}{\coloneqq}

\newcommand{\x}{\mathbf{x}}
\newcommand{\y}{\mathbf{y}}

\begin{document}
\title{Relaxation of first-class constraints and the quantization of gauge theories:\\from ``matter without matter'' to the reappearance of time in quantum gravity}

\author{Roberto Casadio}
\email{roberto.casadio@bo.infn.it}
\affiliation{Dipartimento di Fisica e Astronomia, Universit\`{a} di Bologna,
via Irnerio 46, 40126 Bologna, Italy\\I.N.F.N., Sezione di Bologna, viale B. Pichat 6/2, 40127 Bologna, Italy}

\author{Leonardo Chataignier}
\email{leonardo.chataignier@ehu.eus}
\affiliation{Dipartimento di Fisica e Astronomia, Universit\`{a} di Bologna,
via Irnerio 46, 40126 Bologna, Italy\\I.N.F.N., Sezione di Bologna, viale B. Pichat 6/2, 40127 Bologna, Italy}
\affiliation{Department of Physics and EHU Quantum Center, University of the Basque Country UPV/EHU, Barrio Sarriena s/n, 48940 Leioa, Spain}

\author{Alexander~Yu.~Kamenshchik}
\email{kamenshchik@bo.infn.it}
\affiliation{Dipartimento di Fisica e Astronomia, Universit\`{a} di Bologna,
via Irnerio 46, 40126 Bologna, Italy\\I.N.F.N., Sezione di Bologna, viale B. Pichat 6/2, 40127 Bologna, Italy}

\author{Francisco G. Pedro}
\email{francisco.soares@unibo.it}
\affiliation{Dipartimento di Fisica e Astronomia, Universit\`{a} di Bologna,
via Irnerio 46, 40126 Bologna, Italy\\I.N.F.N., Sezione di Bologna, viale B. Pichat 6/2, 40127 Bologna, Italy}

\author{Alessandro Tronconi}
\email{tronconi@bo.infn.it}
\affiliation{Dipartimento di Fisica e Astronomia, Universit\`{a} di Bologna,
via Irnerio 46, 40126 Bologna, Italy\\I.N.F.N., Sezione di Bologna, viale B. Pichat 6/2, 40127 Bologna, Italy}

\author{Giovanni Venturi}
\email{giovanni.venturi@bo.infn.it}
\affiliation{Dipartimento di Fisica e Astronomia, Universit\`{a} di Bologna,
via Irnerio 46, 40126 Bologna, Italy\\I.N.F.N., Sezione di Bologna, viale B. Pichat 6/2, 40127 Bologna, Italy}

\begin{abstract}\vspace{-0.1cm}
We make a conceptual overview of a particular approach to the initial-value problem in canonical gauge theories. We stress how the first-class phase-space constraints may be relaxed if we interpret them as fixing the values of new degrees of freedom. This idea goes back to Fock and Stueckelberg, leading to restrictions of the gauge symmetry of a theory, and it corresponds, in certain cases, to promoting constants of Nature to physical fields. Recently, different versions of this formulation have gained considerable attention in the literature, with several independent iterations, particularly in classical and quantum descriptions of gravity, cosmology, and electromagnetism. In particular, in the case of canonical quantum gravity, the Fock--Stueckelberg approach is relevant to the so-called problem of time. Our overview recalls and generalizes the work of Fock and Stueckelberg and its physical interpretation with the aim of conceptually unifying the different iterations of the idea that appear in the literature and of motivating further research.
\end{abstract}

\maketitle

\section{Introduction}
Canonical gauge theories are the Hamiltonian description of theories that possess local symmetries (also known as ``gauge symmetries'' or ``gauge redundancies''). The phase-space generators of these local symmetries are associated with a set of initial-value constraints that are of first class, i.e., they Poisson-commute on the phase-space hypersurface defined by all the constraints \cite{Rosenfeld,SalisburyRosenfeld,Dirac1,Dirac2,Dirac3,Bergmann,RBD,Hanson,HT:book}. Yang--Mills theories and electromagnetism are the well-known examples but also general relativity is such a theory, as it is invariant under spacetime diffeomorphisms, which constitute local symmetries (albeit not of the same kind as Yang--Mills theories). In particular, diffeomorphism invariance implies the Hamiltonian vanishes as a constraint \cite{Rosenfeld,Bergmann,DiracGravity,Kiefer:book},
\eq{\label{eq:Hzero}
H = 0 \ ,
}
and this means that not all initial values of fields and canonical momenta are allowed. If the theory is quantized \`{a} la Dirac \cite{Dirac3}, the constraint \eref{Hzero} leads to a time-independent Schr\"odinger equation,
\eq{\label{eq:WDW}
\hat{H}\ket{\Psi} = 0 \ ,
}
also known as the Wheeler-DeWitt equation \cite{Kiefer:book}, which implies quantum states are stationary and the usual notion of time evolution is trivial. This is the so-called problem of time in quantum gravity, and it has inspired many proposed solutions \cite{Anderson:book}. 

One obvious solution is to construct a Hamiltonian that is not constrained. In other words, the Hamiltonian constraint should simply not be satisfied (it should be relaxed). Instead, one considers field configurations or quantum states that do not satisfy Eqs. \eref{Hzero} or \eref{WDW}, so that the Hamiltonian does not vanish, and thus we have the standard unconstrained evolution (given by the time-dependent Schr\"odinger equation). This would describe the ``Schr\"odinger evolution of the Universe'' in the simple setting of a ``minisuperspace'' toy model (i.e., a mechanical model obtained by considering a homogeneous model universe). This trivial solution to the problem of time in quantum cosmology was recently discussed in the article \cite{Kaplan}, and it has also been the subject of many previous works (see, e.g., \cite{HT-CC-1,UnruhUni,UnruhWald,HT-CC-2,CG1,CG2,CG3,GT1,GT2,GT3,GT4,Magueijo1,Magueijo2,Magueijo3}). To the best of our knowledge, the idea has its roots in the works of Fock in 1937 \cite{Fock} and Stueckelberg in 1941 and 1942 \cite{Stueck,Stueck1,Stueck2}, although these articles are seldom cited.

Fock and, more explicitly, Stueckelberg proposed a relaxation of the relativistic particle's mass-shell constraint, which is the particle's version of Eq. \eref{Hzero} (the mass-shell constraint follows from diffeomorphism invariance on the worldline, also referred to as time reparametrization invariance \cite{Chataig:Thesis}). The lack of a mass-shell constraint simply implies that, instead of fixing the particle's (rest) mass to be a certain constant (and thereby restricting the particle's motion to a timelike worldline), its mass should rather be determined by the initial values of the particle's four-momentum, so that different trajectories could yield different masses. Moreover, the trajectories need no longer be timelike, a phenomenon that Stueckelberg used to describe pair creation and annihilation by identifying antiparticles with particles going backwards in Minkowski time.

This idea of Fock and Stueckelberg may also be seen as a passage from the Jacobi action principle (see, e.g., \cite{Lanczos,GrybJ,Chataig:Thesis}), which yields trajectories in configuration space in reparametrization-invariant way (without a preferred time variable), to the usual Hamilton action principle of mechanics, in which there is a preferred time parameter. In the context of general canonical gauge theories, the Fock--Stueckelberg approach is obtained by fixing a gauge condition in the action functional before varying the action, i.e., by imposing the principle of stationary action for a set of gauge-fixed variables. This automatically leads to the ``loss'' of a set of equations of motion (the constraints) which are responsible for the original gauge invariance. In other words, the constraints are relaxed. Notice that this has nothing to do with quantum theory, but rather with defining two different classical theories, as fixing the gauge at the level of the action or the equations of motion generally yields different results.

Stueckelberg's work has inspired an approach to relativistic particle dynamics developed by Horwitz, Piron and others \cite{HPStueck1,HPStueck2,HPStueck3}, which is also predicated on a relaxation of the Hamiltonian constraint. As shown by Stueckelberg, the corresponding quantum theory features a Schr\"odinger equation instead of the usual Klein--Gordon constraint, which is only obtained for energy (mass) eigenstates. Fock and Stueckelberg already noted that this approach corresponds to a modification of the traditional special-relativistic dynamics of the particle. Indeed, the formalism can be understood to be equivalent to promoting the mass to a new degree of freedom (rather than a constant), which is conjugate to the particle's proper time, also elevated to a new degree of freedom. The unconstrained dynamics can then be recast as a new type of constrained (diffeomorphism-invariant) dynamics, where the constraint fixes not the particle's four-momentum, but the initial values of the mass field \cite{Chataig:Thesis}.

The ideas of the recent article \cite{Kaplan} were subsequently developed in the sequel papers \cite{Kaplan1,Kaplan2}, where the authors discuss the relaxation of constraints in field theory, specifically in general relativity and electromagnetism. They argued such a relaxation leads to the development of a well-defined quantization of canonical gauge theories, and they subsequently analyzed the meaning of this relaxation in the classical limit. Hence, their title ``the classical equations of motion of quantized gauge theories.''

In \cite{Kaplan1}, it was noted that, going beyond mechanical (``minisuperspace'') models, the relaxation of the constraints of general relativity leads to solutions to the full set of Einstein equations in the presence of some additional types of matter (dark matter or dark energy). This is simply another case of including new degrees of freedom \`{a} la Fock--Stueckelberg, and, indeed, similar ideas have been repeatedly reported in the literature, where one considered the promotion of constants of Nature to new fields that would be constrained by the value of the Hamiltonian in general relativity (e.g., \cite{Magueijo1,Magueijo2,Magueijo3,Magueijo4,Magueijo5,Magueijo6}), thereby freeing one from the problem of actually solving the Hamiltonian constraint for the original metric and matter variables. A particular case is the elevation of the cosmological constant to a new field, which is not fixed {\it a priori} but determined by the initial values of the matter and geometric fields, in analogy to the relativistic particle's rest mass. This is effectively what one obtains in different approaches to unimodular gravity, which was put forward already by Einstein as early as in 1919 \cite{Einstein}, and later developed in many different ways \cite{HT-CC-1,UnruhUni,UnruhWald,HT-CC-2}.  

Unimodular gravity theories have attracted a considerable amount of attention in recent years. As is well-known, an approach to unimodular gravity can be obtained by the requirement that the theory be invariant with respect to a restricted class of diffeomorphisms, the so-called volume-preserving, area-preserving or divergenceless diffeomorphisms \cite{Arnold,Wolski,Bose,Arakelian,Floratos,Bars,Wit,closed,Kugo}; i.e., by the diffeomorphisms generated by the divergenceless transverse vector fields. In such a formalism, the Hamiltonian constraint \eref{Hzero} is not satisfied at all points, but the value of the Hamiltonian [the left-hand side of the constraint equation \eref{Hzero}] is related to an integration constant that plays the role of the cosmological constant. As in the case of the Fock--Stueckelberg formalism for the relativistic particle, the approaches to unimodular gravity can be seen as particular modifications of general relativity, whereby one adds new degrees of freedom and modifies the local symmetry group, which leads to a relaxation of the original constraints.

Evidently, this type of modification is not restricted to gravitational theories. In principle, one can apply the Fock--Stueckelberg reasoning to any canonical gauge theory. We will variously refer to such treatments of canonical gauge theories as the Fock--Stueckelberg theory, approach, mechanism, or method. One simple example can be done for electromagnetism, as discussed by the authors of \cite{Kaplan} and \cite{Kaplan1} in the article \cite{Kaplan2}, where they relaxed the standard vacuum Gauss-law constraint, ${\rm div} \vec{E}=0$, replacing it by
\begin{equation}
{\rm div} \vec{E} = \rho \ ,
\end{equation}  
where $\vec{E}$ is the electric field and $\rho$ encodes the now relaxed (possibly nonzero) value of the constraint. It can be interpreted as an electric charge density. With this relaxation, they observed the ``emergence of charge without matter.'' This is, again, an iteration of the Fock--Stueckelberg approach and, as a matter of fact, the theory of electrodynamics without the Gauss-law constraint was already considered by Dirac in 1951 \cite{Dirac}. Dirac showed in \cite{Dirac} that, by discarding the Gauss-law constraint, one acquires an additional degree of freedom, which can be identified with the electron. The ideas of Dirac's new electrodynamics were further developed by many authors (see, e.g., \cite{post,post1,post11,post2,post3}). 

Although the formalism proposed by the authors of \cite{Kaplan,Kaplan1,Kaplan2} may certainly lead to a well-defined quantum theory of interest, it is important to emphasize that the relaxation of constraints is by no means a mandatory step in the quantization of gauge theories. This is evidenced by the success of quantum electrodynamics, not to mention the standard model of particle physics, and by the powerful formalism of quantization \`{a} la Batalin--Vilkovisky (BV), Batalin--Fradkin--Vilkovisky (BFV), and Becchi--Rouet--Stora--Tyutin (BRST) (see, e.g., \cite{HT:book}). In this way, although it may be easier to quantize a theory without (some of the) constraints, the relaxation of constraints is not a prerequisite for a well-defined quantum theory (at least given our current knowledge of quantum foundations). Rather, as was already mentioned, the relaxation of constraints defines a different {\it classical} theory. Both theories, constrained and unconstrained, may be quantized and each may lead to different phenomenology due to the presence or absence of new degrees of freedom (e.g., presence or absence of a mass field in the traditional and Fock--Stueckelberg treatments of the relativistic particle). This is a central point of our analysis.

Furthermore, it is true that there is a distinction between Yang--Mills symmetries and diffeomorphism invariance but this need not imply that the standard treatment of the constraints in general relativity will lead to a pathological quantum theory. The Yang--Mills symmetries are internal, as they do not concern a change of spacetime point (the argument of fields), whereas diffeomorphisms are external as they ``move points.'' However, by taking an active view of diffeomorphisms, in which the original and transformed fields may differ in functional form but not in argument, this distinction seems less crucial. In fact, both types of local symmetries can be treated in formally the same way when analyzing the invariance of the action functional, provided one adds suitable boundary terms \cite{Vergara}. It thus seems premature to {\it a priori} dismiss the standard quantization of general relativity's constraints. The task of finding such a (possibly ultraviolet complete) quantization that is non-perturbative is surely daunting but not necessarily impossible. In this way, the Fock--Stueckelberg approach is not inevitable in quantum gravity and cosmology.

Evidently, however, the necessity of going beyond the standard models of particle physics and cosmology may motivate us to consider the elegant method of Fock and Stueckelberg as a means of accounting for new physics. Although this was indeed the spirit of the articles \cite{Kaplan,Kaplan1,Kaplan2}, and of the many other preceding articles cited above, an overview with an emphasis on the physical meaning of the Fock--Stueckelberg theory is long overdue, due to the frequent appearance of versions of this approach in the literature. In the present article, we propose such an overview, which not only recalls the original works of Fock and Stueckelberg, but also offers a generalization that encompasses different approaches in the literature.

The title of the present article mentions the relaxation of first-class constraints and the quantization of gauge theories. What is their relation? As discussed above, and as will be argued throughout, this relaxation is not needed to achieve a consistent and general quantization of gauge theories, despite what some of the literature may imply. We will show below that the relaxation of constraints à la Fock and Stueckelberg is a mechanism to define new classical theories, which can then be quantized using standard techniques. This also has implications for the problem of time in quantum gravity and its solution.

The article is divided into five sections and an appendix. In Sec. \ref{sec:gauge-theory}, we quickly review the fundamentals of canonical gauge theories (with further details in Appendix \ref{app:noether}), in order to subsequently present the Fock--Stueckelberg approach to general gauge theories in Sec. \ref{sec:FS}. This section is then subdivided into several particular cases and examples of interest. In Sec. \ref{sec:constants}, we discuss how promoting constants of Nature to spacetime fields can be seen as an equivalent approach to the Fock--Stueckelberg theory under certain conditions. The constraint relaxation for ``partially Abelianized'' systems is then presented in Sec. \ref{sec:subalgebra}, and subsequently we turn our attention to mechanical examples in Secs. \ref{sec:mechsimple} and \ref{sec:mech}, where we examine the case of time reparametrization invariance and relativistic particles. The Fock--Stueckelberg approach to electromagnetism is presented in Sec. \ref{sec:DiracEM}, whereas the application to general relativity is featured in Sec. \ref{sec:gug-0}. Section \ref{sec:reappear} discusses the implications of the Fock--Stueckelberg method to the problem of time in quantum gravity, and we finish with our conclusions in Sec. \ref{sec:conclusions}.

\section{\label{sec:gauge-theory}Canonical gauge theories in a nutshell}
We begin with a brief review of canonical gauge theories before we discuss a general presentation of what we call the Fock--Stueckelberg approach. Their original work on the relativistic particle will be discussed as a particular example of the general framework. We focus on the canonical description of gauge theories because the relation between gauge symmetries and initial-value constraints is most clear in this formulation. As we will see, this leads to a clear implementation of the Fock--Stueckelberg mechanism and its connection to a relaxation of first-class constraints, as well as to the appearance of new degrees of freedom (such as ``charge without charge'' and ``matter without matter'').

\subsection{\label{sec:gauge-transf}Constraint algebra and gauge transformations}
Let us consider a set of fields $z^{m}(\tau,\mathbf{x})$ ($m=1,\ldots,Z$) in $d+1$ spacetime dimensions. We omit the spatial dependence by adopting DeWitt's compact notation \cite{DeWittFields} in which abstract indices include not only the discrete values of indices such as $m$, but also the spatial dependence of the fields. For example, $z^{m}(\tau,\mathbf{x}) \equiv z^{\mu}(\tau)\equiv z^{\mu}$,\footnote{We use the sign $\equiv$ to denote equivalent notations. It is also used when two functions are identically equal (i.e., equal at every point). We will use the sign $\Def$ to denote the definition of a quantity.} where $\mu$ is the abstract index that includes both $m$ and $\mathbf{x}$. Repeated indices are then taken to imply a tacit summation over the discrete values of the original indices (e.g., over $m = 1,\ldots, Z$) together with a spatial integration. In what follows, as a matter of notational convenience, we also write $\mu = 1,\ldots,Z$ and similar expressions for other abstract indices with the understanding that this slight abuse of notation only specifies the discrete values covered by the abstract index. In the case $d=0$ (mechanics), there is of course no spatial dependence and no spatial integrations. We assume that the fields $z^{\mu}$ are bosonic for simplicity. The field-space spanned by $z^{\mu}$ will be referred to as the phase space.

A canonical gauge theory can be defined by the action functional\footnote{\label{foot:functionals}We use square brackets $[\cdot]$ to denote that a quantity is a functional of its arguments, and round brackets $(\cdot)$ to denote an ordinary function of the arguments. A combination of brackets $(\cdot\,;\,\cdot]$ denotes a quantity that is a function of some of its arguments and a functional of the others. When no confusion is possible, we may omit the arguments. For example, if $d>0$, we may have $H_0(\tau;z] = \int\D^dx \mathscr{H}_0(\tau;z(\tau,\mathbf{x}),\partial z(\tau,\x),\ldots,\partial^n z(\tau,\x))$ with some smooth function $\mathscr{H}_0$ and $\partial^n$ being the $n$-th spatial derivative. A function can also be written as a functional: $f(z(\tau;\mathbf{x})) = \int\D^dy f(z(\tau;\mathbf{y}))\delta(\mathbf{y},\mathbf{x})\equiv f[z]$.}
\eq{\label{eq:action}
S[z,\lambda] = \int_{\tau_0}^{\tau_1}\D\tau \left\{\mathcal{A}_{\mu}[z]\dot{z}^{\mu}-H_0(\tau;z]-\lambda^a\phi_a(\tau;z]\right\} \,,
}
where a dot indicates a derivative with respect to time $\tau$, $\A_{\mu}$ is the symplectic potential, and $\lambda^a$ are Lagrange multipliers ($a=1,\ldots,N\leq Z$). The field equations of $z^{\mu}$ are\footnote{\label{foot:fder}The derivatives with respect to $z^{\mu}$ are taken in the functional sense if $d>0$. We recall that, for a local functional $\mathcal{F}$ of the field $z(\tau,\mathbf{x})$, which is written as $\mathcal{F}(\tau,\x;z] = \int\D^d y F\left(\tau,\x,\y;z(\tau,\y), \partial z(\tau,\y),\ldots,\partial^n z(\tau,\y)\right)$ with $F$ being a scalar density in $\mathbf{y}$, the functional derivative $\delta \mathcal{F}/\delta z(\tau,\x)$ of $\mathcal{F}$ relative to a variation $z(\tau,\x)\mapsto z(\tau,\x)+\varepsilon\eta(\tau,\x)$ is obtained from $\delta \mathcal{F} = \varepsilon\left.\Del{}{\varepsilon}\right|_{\varepsilon=0}\mathcal{F}(\tau,\x;z+\varepsilon \eta] =: \varepsilon\int\D^dy\ \frac{\delta \mathcal{F}}{\delta z(\tau,\y)}\eta(\tau,\y)$. This leads to $\delta z(\tau,\x) = \varepsilon \eta(\tau,\x)$, as it should.}
\eq{\label{eq:eom1}
\Omega_{\mu\nu}\dot{z}^{\nu} = \del{H}{z^{\mu}} \ ,
}
where
\eq{
\Omega_{\mu\nu} &:= \del{\A_{\nu}}{z^{\mu}}-\del{\A_{\mu}}{z^{\nu}} \ , \\
H &:= H_0+\lambda^a\phi_a
}
are the symplectic form and the total Hamiltonian, respectively. We assume that $\Omega_{\mu\nu}$ is invertible with $\Omega^{\mu\nu}\Omega_{\nu\rho} = \delta^{\mu}_{\rho}$.\footnote{If $d>0$, the delta symbol corresponds to a multiplication of a Kronecker delta for the discrete variables and a Dirac delta distribution for the continuous spatial coordinates.} Then, Eq. \eref{eom1} can be rewritten as
\eq{\label{eq:eom}
\dot{z}^{\mu} = \Omega^{\mu\nu}\del{H}{z^{\nu}} \ .
}
This motivates the definition of the Poisson bracket of two function(al)s $f$ and $g$ as
\eq{
\{f,g\} := \del{f}{z^{\mu}}\Omega^{\mu\nu}\del{g}{z^{\nu}} \ .
}
In this way, the field equations associated with $z^{\mu}$ can be written compactly as $\dot{z}^{\mu} = \{z^{\mu},H\}$. On the other hand, the field equations associated with the multipliers $\lambda^a$ are constraints on the fields:
\eq{\label{eq:constraints}
\phi_a(\tau;z] = 0 \ .
}
We can now make a series of simplifying assumptions that will be sufficient for our purposes. We assume that $\phi_a$ are functions of the fields and their spatial derivatives, $\phi_a\equiv\phi_a(z,\partial z,\ldots,\partial^n z)$ with $n$ finite, and they do not depend explicitly on time,
\eq{
\del{\phi_a}{\tau} = 0 \ .
}
Moreover, the $\phi_a$ functions are all functionally independent and not identically zero, so there is no constraint that can be reduced to another (e.g., $\phi_2 = \phi_1^2$). The constraint equations \eref{constraints} together with their first $p$ derivatives ($\partial\phi_a = 0,\ldots,\partial^p\phi_a = 0$) are also assumed to define a hypersurface $\Sigma$ on the space of the $\mathtt{Z} = (z,\partial z,\ldots,\partial^{n+p}z)$ values, which can be taken to be independent at a fixed arbitrary $\x$. If $\Phi_\alpha = 0$ are the independent equations among $(\phi_a = 0,\ldots,\partial^p\phi_a = 0)$, then in open regions covering $\Sigma$, we assume that $\Phi_{\alpha}$ can be regarded as coordinates in a neighborhood of $\Sigma$; i.e., we can define local coordinates $\mathtt{Z}'(\mathtt{Z}) = (\tilde{\mathtt{Z}}(\mathtt{Z}),\Phi(\mathtt{Z}))$. Then, if $f'(\mathtt{Z}') \Def f(\mathtt{Z}(\mathtt{Z}'))$ is such that $f'(\tilde{\mathtt{Z}},\Phi=0)=0$, we obtain\footnote{See Theorem 1.1, Appendix 1.A and Theorem 12.4 in \cite{HT:book}. In Eq.~\eref{KT-0}, the derivative with respect to $\ell$ is an ordinary partial derivative.}
\eq{\label{eq:KT-0}
f'(\tilde{\mathtt{Z}},\Phi) &= \int_0^1\D\ell\ \del{}{\ell}f'(\tilde{\mathtt{Z}},\ell\Phi)\\
&= \Phi_\alpha\int_0^1\D\ell\ \frac{1}{\ell}\del{}{\Phi_\alpha}f'(\tilde{\mathtt{Z}},\ell\Phi)\\
&=:V^\alpha(\tilde{\mathtt{Z}},\Phi)\Phi_\alpha \ .
}
Thus, if a function $f(\mathtt{Z})$ vanishes when the constraints are satisfied, it vanishes on $\Sigma$, and it can be written as a combination of $(\phi_a,\ldots,\partial^p\phi_a)$ with some finite $p$ and with coefficients that are locally defined by Eq. \eref{KT-0}. That this is achievable without the use of boundary conditions (but only with the use of $\Phi_\alpha = 0$) corresponds to the assumption of `local completeness' of the constraint functions, which, just as the fact that $\Phi_\alpha$ serve as local coordinates, constitutes a regularity condition to be imposed on the constraints.

We use Dirac's `weak equality' sign $\approx$ to denote identities that hold when the constraints are satisfied \cite{Dirac2}.\footnote{Identities that hold when (all) the field equations are satisfied can be called `on-shell identities'. Since every solution to the field equations must satisfy Eqs. \eref{constraints}, then on-shell identities are also weak equalities. In a slight abuse of terminology, we use both weak equalities and on-shell identities as synonyms.} In particular, $\phi_a\approx0$. The constraint equations \eref{constraints} must hold at every instant of time, so that $\dot{\phi}_a\approx0$, otherwise the theory is not consistent. The requirement that $\dot{\phi}_a\approx0$ may lead to new constraints $C_a$ that are functionally independent from $\phi_a$. In the standard Rosenfeld--Bergmann--Dirac analysis \cite{Rosenfeld,SalisburyRosenfeld,Dirac3,RBD,HT:book,Bergmann}, the $\phi_a$ are called `primary' constraints, as they follow from the structure of the action \eref{action}, whereas $C_a$ are called `secondary' as they follow from the application of the field equations.\footnote{Repeated application of the field equations might lead to `tertiary' constraints and so forth, but all these constraints can be collectively called secondary.} For our purposes, it is sufficient to consider that only the primaries $\phi_a$ are present; i.e., the conditions $\dot{\phi}_a\approx0$ are satisfied without leading to new (secondary) constraints. In this case, if we demand that $\dot{\phi}_a\approx 0$ is satisfied for every arbitrary value of the multipliers $\lambda^a$, then $\phi_a$ and $H_0$ must obey a first-class algebra (one that Poisson-commutes if the constraints hold). Due to Eq. \eref{KT-0}, this implies 
\eq{\label{eq:first-class}
\{\phi_a,\phi_b\} &= \mathcal{F}_{ab}^c\phi_c \approx 0 \ ,\\
\{\phi_a,H_0\} &= \mathcal{V}_a^b\phi_b \approx 0 \ ,
}
with structure function(al)s $\mathcal{F}_{ab}^c[z] \equiv \mathcal{F}_{ab}^c$ and $\mathcal{V}_a^b[z]\equiv \mathcal{V}_a^b$.\footnote{For brevity, we refer to $\mathcal{F}_{ab}^c$ and $\mathcal{V}_{a}^b$ as structure functions, even though they may be functionals in general.} With this, $\phi_a\approx0$ can be seen as restrictions on the allowed initial values of the $z$ fields, and the Lagrange multipliers $\lambda^a$ are not determined by any field equation, be it Eqs. \eref{eom} or Eqs. \eref{constraints}. On the other hand, the Eqs. \eref{eom} depend on $\lambda^a$. In this way, solutions to the field equations will depend on arbitrary functions $\lambda^a$. This is a defining feature of a gauge theory.

What is the significance of these arbitrary functions? They can be thought of as determining a ``generalized reference frame'' or gauge \cite{HT:book,Chataig:Thesis}. The fact that the theory does not determine $\lambda^a$ implies that no frame is preferred. A change of frame $\lambda^a\to\lambda^a+\delta\lambda^a$ is a gauge transformation, which also changes the solution $z^{\mu}\to z^{\mu}+\delta z^{\mu}$ due to the dependence of Eqs. \eref{eom} on $\lambda^a$.

Is any such transformation allowed? No. If no preferred frame is to be selected, then the field equations must have the same functional form in all frames. This means that $z^{\mu}\to z^{\mu}+\delta z^{\mu}$, $\lambda^a\to\lambda^a+\delta\lambda^a$ should be a symmetry. Then, the Noether charge $G$ must satisfy (see Appendix \ref{app:noether}):
\eq{\label{eq:Noether}
\left(\Omega_{\mu\nu}\dot{z}^{\nu}-\del{H}{z^{\mu}}\right)\delta z^{\mu}-\phi_a\delta\lambda^a+\Del{G}{\tau} = 0 \ .
}
This is an `off-shell identity'; i.e., it holds regardless of whether the field equations are satisfied. If the field equations hold, we obtain the `on-shell identity' $\D G/\D\tau = 0$, which is the conservation of the Noether charge. As Eq. \eref{Noether} holds for arbitrary off-shell paths, we can take $z^{\mu}$ and $\dot{z}^{\mu}$ (as well as $\lambda^a,\dot{\lambda}^a$) to be independent quantities, and we can set their coefficients to zero separately. The coefficients of $\dot{z}^{\mu}$ and $\dot{\lambda}^a$ lead to
\eq{\label{eq:Noether2}
\delta{z}^{\mu} &= \{z^{\mu},G\} \ , \\
\del{G}{\lambda^a} &= 0 \ ,
}
and this implies that the Noether charge generates a canonical transformation that is independent of the (arbitrary) multipliers. The remaining condition from Eq. \eref{Noether} is
\eq{\label{eq:Noether3}
\phi_a\delta\lambda^a = \del{G}{\tau}+\{G,H\} \ ,
}
which, due to the arbitrariness of $\delta\lambda^a$ and the fact that $\phi_a\neq0$ off shell, implies that $G$ must depend on $N$ arbitrary functions of time $\varepsilon^a$. We can thus set $G = \varepsilon^aG_a$ and use the independence of $\varepsilon^a$ and their time derivatives to find that $G_a$ must be a linear combination of the primary constraints. Without loss of generality, we can set $G_a = \phi_a$, such that Eq. \eref{Noether3} together with Eqs. \eref{first-class} leads to the transformation law for the multipliers \cite{HT:book,Pons}
\eq{\label{eq:delta-lambda}
\delta\lambda^a = \dot{\varepsilon}^a+\varepsilon^b\mathcal{V}_b^a+\varepsilon^b\lambda^c\mathcal{F}_{bc}^a \ .
}
Notice that, on shell, the quantities $\varepsilon^a$ may also depend on the fields $z^{\mu}$, so that we can identify $\dot{\varepsilon}^a = \partial_\tau\varepsilon^a+\{\varepsilon^a,H\}$. Thus, a gauge transformation is a (possibly field-dependent, on-shell) canonical transformation of the $z$ fields generated by the constraints $\phi_a$ and accompanied by the transformation \eref{delta-lambda} of the multipliers (change of generalized reference frame), described by arbitrary functions $\varepsilon^a$. It is a local (i.e., spacetime-dependent) symmetry transformation, with Noether charge $G = \varepsilon^a\phi_a$.

The on-shell (with possibly field-dependent $\varepsilon^a$) gauge transformation of a general quantity $\mathcal{I}$ reads:
\eq{\label{eq:general-gauge-transf}
\delta\mathcal{I} &= \{\mathcal{I},G\}+\del{\mathcal{I}}{\lambda^a}\delta\lambda^a\\
&\approx \varepsilon^a\left(\{\mathcal{I},\phi_a\}+\mathcal{V}_a^b\del{\mathcal{I}}{\lambda^b}+\mathcal{F}_{ac}^b\lambda^c\del{\mathcal{I}}{\lambda^b}\right)+\del{\mathcal{I}}{\lambda^a}\dot{\varepsilon}^a\,.
}
If $\mathcal{I}$ is gauge invariant, it does not change under arbitrary gauge transformations, i.e., $\delta\mathcal{I}\approx0$ for any choice of $\varepsilon^a$. Using the independence of $\varepsilon^a$ and its derivatives implied by the arbitrariness of $\varepsilon^a$, we see that Eq. \eref{general-gauge-transf} together with $\delta\mathcal{I}\approx0$ implies the gauge invariance conditions:
\eq{\label{eq:gauge-invariants}
\del{\mathcal{I}}{\lambda^a} \approx 0 \ , \ \{\mathcal{I},\phi_a\} \approx 0 \ .
}
A gauge-invariant quantity can be seen as one that is the same in every generalized reference frame. In particular, the constraint functions $\phi_a$ are gauge invariant, $\delta\phi_a\approx0$, and thus their functional form does not change under a gauge transformation.

\subsection{\label{sec:gf}Gauge fixing}
A choice of generalized reference frame consists in a fixation of $\lambda^a$, which corresponds to a choice of gauge or gauge fixing. This can be done by imposing some \emph{ad hoc} constraints (``gauge conditions'') \cite{Hanson,HT:book},
\eq{\label{eq:gauge-condition}
\chi^a(\tau;z,\lambda] \approx 0 \ ,
}
which must hold at all times. Admissible choices must be accessible; i.e., for every arbitrary value of $\lambda^a$, there must exist a gauge transformation with some function $\varepsilon^a$ that takes the arbitrary initial frame to chosen one. If there exists a gauge transformation with some functions $\varepsilon^a\equiv\varepsilon^a_{\star}$ that preserves Eq. \eref{gauge-condition}, then this transformation is called a `residual gauge transformation' (generated by $G_{\star}=\varepsilon^a_{\star}\phi_a$), and the gauge-fixing conditions are said to be `partial' or to `partially fix the gauge'. If there are no residual gauge transformations, the gauge-fixing conditions are said to be `complete' or to `fix the gauge completely'.

The imposition of the extra constraints \eref{gauge-condition} occurs at level of the field equations (i.e., on shell). Indeed, let us consider a complete gauge fixing for simplicity, and let the fixated $\lambda^a$ that results from Eq. \eref{gauge-condition} be $\lambda^a\approx\Lambda^a(\tau;z]$. Then the gauge-fixed field equations are
\eq{\label{eq:gf-eom-0}
\dot{z}^{\mu} &\approx \{z^{\mu},H_0\}+\Lambda^a(\tau;z]\{z^{\mu},\phi_a\} \ ,\\
\phi_a &\approx 0 \ , \ \chi^a(\tau;z,\lambda] \approx 0 \ ,
}
which can then be subsequently solved for $z^{\mu}$. 

There are different ways in which one can consistently implement the gauge fixing at the level of the action.\footnote{A powerful method is related to the BRST cohomology \cite{HT:book}, which we do not cover here.} A simple approach is to introduce new Lagrange multipliers $w_a$ to modify the action \eref{action} in order to introduce the new gauge-fixing constraints:\footnote{The term $(\chi^b)^p$ is the $p$-th power of the gauge condition for each value of $b$ (no summation over $b$ implied). In the term $w_b(\chi^b)^p$, there is one tacit summation over $b$ together with a spatial integration if $d>0$.}
\eq{\label{eq:gf-action}
S_{\rm gf} \Def \int_{\tau_0}^{\tau_1}\!\D\tau\! \left[\mathcal{A}_{\mu}\dot{z}^{\mu}-H_0-\lambda^a\phi_a-w_b(\chi^b)^p\right] \! ,
}
where $p\geq1$ is a natural number. The new gauge-fixed action \eref{gf-action} leads to the field equations:
\eq{\label{eq:gf-eom}
\dot{z}^{\mu} &= \{z^{\mu},H_0\}+\{z^{\mu},\lambda^a\phi_a\}+\{z^{\mu},w_b(\chi^b)^p\} \ , \\
0&\approx \phi_a+w_b\del{}{\lambda^a}(\chi^b)^p =: \phi_a+\tilde{w}_a \ , \\
0&\approx (\chi^b)^p \ .
}
Our goal is to show that these equations form a consistent system that can be made equivalent to Eqs. \eref{gf-eom-0}. The last two sets of equations in \eref{gf-eom} are the new constraints, which must be valid at all times if the dynamics derived from the action \eref{gf-action} is consistent. In this context, we use the sign $\approx$ to indicate equalities on the hypersurface defined by the new constraints. First, notice that the constraints $(\chi^b)^p\approx0$ are equivalent to $\chi^b\approx 0$. Furthermore, in order to recover the original gauge-fixed field equations, we must show that the $w$ multipliers are spurious in the sense that they disappear from Eqs. \eref{gf-eom}. Evidently, this can be achieved if we can fix $w_b \equiv0$ on shell. For $w_b\neq0$, if the last term in the field equations for $z^{\mu}$ is to vanish when $\chi^b\approx0$, then either $p\geq2$ or, if $p=1$, $\chi^b$ must Poisson-commute with all the $z$ fields and, in particular, with $\phi_a$. Furthermore, in order to recover the primary constraints $\phi_a \approx 0$, then the equations in the second line of Eqs. \eref{gf-eom} should not depend on $w_b\neq0$, which implies that $p\geq2$ or, if $p=1$, $\chi^b$ must not depend on the multipliers. We thus see that the case $p=1$ with $w_b\neq0$ is inconsistent: it implies that $\chi^b$ must be gauge invariant [cf. Eqs. \eref{gauge-invariants}], and thus they are not well-defined gauge-fixing conditions, as every gauge transformation preserves them. In this way, we must have $p\geq2$ if $w_b\neq0$ on shell, which leads us back to Eqs. \eref{gf-eom-0} (with $\lambda^a \approx \Lambda^a(\tau;z]$ following from $\chi^a\approx0$, as before).

A useful particular case is that in which the gauge-fixing conditions only depend on the $z$ fields, $\chi^a(\tau;z,\lambda]\equiv\chi^a(\tau;z]$. These are called `canonical gauge-fixing conditions' or simply `canonical gauges'  \cite{Hanson,HT:book}. Then, taking $p=2$ without loss of generality, the conservation equation
\eq{\label{eq:gf-stability}
0 \approx \del{\chi^a}{\tau} +\{\chi^a,H_0\}+\lambda^b\{\chi^a,\phi_b\}
}
leads to a fixation of the $\lambda$ fields if the system of constraints $(\phi_a,\chi^b)$ is not of first class (it is said to be of second class). In particular, we must have the Faddeev--Popov determinant condition \cite{HT:book,FP}
\eq{\label{eq:FP-det}
\det\{\chi^a,\phi_b\} \neq 0 \ .
}
Typically, this condition holds only locally in field space (and thus no canonical gauge fixing condition is valid globally). With this, one can form the inverse matrix $(\Delta^{-1})^{a}_b$, such that $(\Delta^{-1})^{a}_b\{\chi^b,\phi_c\} = \delta^a_c$. The solution for the $\lambda$ fields obtained from Eq. \eref{gf-stability} then reads
\eq{\label{eq:fix-mult}
\lambda^a \approx -\left(\Delta^{-1}\right)^{a}_b\left(\del{\chi^b}{\tau}+\{\chi^b,H_0\}\right) \equiv \Lambda^a(\tau;z] \ ,
}
and the field equations \eref{gf-eom} again coincide with Eqs. \eref{gf-eom-0}.

\section{\label{sec:FS}To gauge or not to gauge? The approach of Fock and Stueckelberg}
\subsection{\label{sec:FS-gen}General formalism}
We have seen that the gauge-fixed action \eref{gf-action} with $p=1$ and $w_b\neq0$ on shell is inconsistent in the sense that it does not yield the gauge-fixed equations of motion \eref{gf-eom-0} for well-defined gauge-fixing conditions, and therefore it is not compatible with the dynamics of the gauge system. However, it is of course legitimate to consider the $p=1$ action
\eq{\label{eq:FS-action}
S_{\rm FS} \Def \int_{\tau_0}^{\tau_1}\!\D\tau\! \left[\mathcal{A}_{\mu}\dot{z}^{\mu}-H_0-\lambda^a\phi_a-w_b\chi^b\right]
}
as defining a new theory, which we call the Fock--Stueckelberg theory. Thus, the Fock--Stueckelberg approach consists in defining a new theory from a ``parent gauge theory.''

What are the properties of the Fock--Stueckelberg theory? The fields to be varied in the action \eref{FS-action} are $z^{\mu}, \lambda^a$ and $w_b$. Thus, we notice that we have new degrees of freedom with respect to the parent theory: the $w_b$ fields. In the action given in Eq. \eref{gf-action} with $p\geq2$, these degrees of freedom are spurious as they do not affect the field equations. This is not necessarily so in the Fock--Stueckelberg theory, where the field equations are given by Eqs. \eref{gf-eom} with $p=1$.

As we have seen in Sec. \ref{sec:gf}, if $\chi^a$ are well-defined gauge-fixing conditions in the parent theory, then they cannot be independent of $\lambda^a$ and simultaneously Poisson-commute with the $z$ fields, as they cannot be gauge invariant. Let us now analyze the case in which $\chi^a$ are (not) canonical gauges of the parent theory.

\subsubsection{Canonical gauges of the parent theory}
If $\partial\chi^b/\partial\lambda^a$ vanishes identically, then we have canonical gauge conditions. In this case, $\tilde{w}_a=0$ [cf. Eqs. \eref{gf-eom} with $p=1$], and the constraints of the Fock--Stueckelberg theory coincide with the first-class constraints $\phi_a$ and the gauge conditions $\chi^b$ of the parent theory. These constraints do not fix the $w$ multipliers. However, the conservation of $\phi_a\approx0$ in time with $\tilde{w}_a =0$ and $p=1$ implies that
\eq{
0\approx\dot{\phi}_a \approx w_b\{\phi_a,\chi^b\} \ ;
}
i.e., $w_a$ must be zero on shell [as the canonical gauges $\chi^b$ cannot Poisson-commute with all the $\phi$ constraints, cf. Eq. \eref{FP-det}].\footnote{This is true for a complete canonical gauge fixing. For a partial canonical gauge fixing, we can introduce $N_u<N$ canonical gauges $\chi^{a_u}$ ($a_u = 1,\ldots,N_u$), which Poisson-commute with $\phi_{a>N_u}$ but must obey $\det\{\chi^{a_u},\phi_{b_u}\}\neq0$ ($a_u,b_u = 1,\ldots,N_u$) in order to fix $N_u$ of the $\lambda$ multipliers via the immediate analogue of Eq. \eref{gf-stability}. Thus, in the Fock--Stueckelberg theory, we introduce $N_u$ multipliers $w_{a_u}$. Then, $\dot{\phi}_{a>N_u}\approx0$ is an on-shell identity, whereas $\dot{\phi}_{a_u}\approx w_{b_u}\{\phi_{a_u},\chi^{b_u}\}$ again implies that $w_{b_u}\approx0$.} In this way, the Fock--Stueckelberg theory coincides with the parent theory in canonical gauges, and thus it does not lead to new physical effects.

\subsubsection{\label{sec:noncanon}Noncanonical gauges and the relaxation of the parent theory's constraints}
If $\partial\chi^b/\partial\lambda^a\neq0$, then the constraint $\phi_a+\tilde{w}_a\approx0$ [cf. Eqs. \eref{gf-eom} with $p=1$] can be seen as a fixation of $\tilde{w}_a$ rather than a restriction on the possible initial values of the $z$ fields. The original first-class constraints of the parent theory are thus relaxed: we can now allow arbitrary initial values of the $z$ fields, as this will only change the corresponding $\tilde{w}_a$ field. The values of $z^{\mu}$ for which $\phi_a\approx0$ are recovered only as a particular case, in which $\tilde{w}_a$ is correspondingly fixed to be zero on shell.\footnote{If $\partial\chi^b/\partial\lambda^a\neq0$ are the coefficients of an invertible matrix, then we see from Eqs. \eref{gf-eom} with $p=1$ that $\tilde{w}_a\approx0$ implies $w_a\approx0$, in which case the Fock--Stueckelberg theory becomes again equivalent to the parent theory. On the other hand, if $\partial\chi^b/\partial\lambda^a\neq0$ form a noninvertible matrix, then $\tilde{w}_a\approx0$ only implies that $w_b$ is a possibly nontrivial null eigenvector, for which the Fock--Stueckelberg theory need not coincide with its parent.} Therefore, the inclusion of the new degrees of freedom $w_a$ effectively removes the first-class constraints (and thus the gauge symmetry) of the parent theory. This is the key feature of the Fock--Stueckelberg approach. Evidently, the fixation of $\tilde{w}_a$ (and, correspondingly, of $w_a$) via $\phi_a+\tilde{w}_a\approx0$ and of the $\lambda$ multipliers via $\chi^a\approx0$ must hold at all times. This will be illustrated in the examples below.

For example, a case of interest in applications of the Fock--Stueckelberg approach is given by the simple noncanonical gauge-fixing conditions
\eq{\label{eq:FS-gf-main}
\chi^a(\tau;z,\lambda] &\Def (\lambda^{a_u}-\Lambda^{a_u}(\tau;z])\delta_{a_u}^a \ ,\\
 a_u &= 1,\ldots,N_u\leq N \ ,
}
which fix the first $N_u\leq N$ gauge multipliers (thus, it may be a partial gauge fixing). In this case, the Fock--Stueckelberg constraints read
\eq{\label{eq:FS-constraints}
\chi^{a_u}&\approx0 \ , \\
\phi_{a_u}+w_{a_u}&\approx0\ , \\
\phi_{a>N_u}&\approx0 \ ,
}
and the field equations for $z^{\mu}$ [cf. Eqs. \eref{gf-eom}] can be rewritten as
\eq{\label{eq:FS-eom}
\dot{z}^{\mu} &= \{z^{\mu},H_0+\lambda^a\phi_a+w_{a_u}\chi^{a_u}\}\\
&\approx \{z^{\mu},H_0\}+\Lambda^{a_u}\{z^{\mu},\phi_{a_u}+w_{a_u}\}\\
&\ \ \ +\{z^{\mu},\lambda^{a>N_u}\phi_{a>N_u}\}-\{z^{\mu},w_{a_u}\Lambda^{a_u}\}\\
&\approx \{z^{\mu},H_0+\Lambda^{a_u}\phi_{a_u}+\lambda^{a>N_u}\phi_{a>N_u}\} \ ,
}
where we used the constraints given in Eqs. \eref{FS-constraints}. Notice that $w_{a_u}$ have disappeared from these equations, and their only effect is to relax the constraints $\phi_{a_u}$, as discussed above. Thus, the field equations \eref{FS-eom} together with the constraints given in Eqs. \eref{FS-constraints} are equivalent on shell to the equations derived from the action functional
\eq{\label{eq:FS-action-1}
\!\tilde{S}_{\rm FS} \Def \int_{\tau_0}^{\tau_1}\!\D\tau\! \left[\mathcal{A}_{\mu}\dot{z}^{\mu}-H_0-\Lambda^{a_u}\phi_{a_u}-\lambda^{a>N_u}\phi_{a>N_u}\right] \,,
}
which is simply obtained from the action \eref{action} by imposing $\lambda^{a_u} = \Lambda^{a_u}(\tau;z]$ before variation. From the point of view of the parent theory, this is an erroneous gauge-fixing procedure, whereas it is a mechanism to relax the constraints and introduce new degrees of freedom in the Fock--Stueckelberg theory.\footnote{As in the parent gauge theory, the Rosenfeld--Bergmann--Dirac algorithm \cite{Rosenfeld,SalisburyRosenfeld,Dirac3,RBD,HT:book,Bergmann} that ensures the theory's consistency by verifying that the constraints hold at all times should also be applied to the Fock--Stueckelberg theory. This means that we must require that the time derivatives of the constraints derived, for example, from the action given in Eq. \eref{FS-action-1} vanish on shell. This may lead to a fixation of multipliers $\lambda$, to redundant constraints or to new constraints. In case new constraints appear, we must again require they hold at all times.}

Let us now consider a series of illustrating examples and particular cases of the Fock--Stueckelberg theory.

\subsection{\label{sec:constants}Constants of Nature as spacetime fields}
Before we analyze more concrete examples, it is interesting to note that the Fock--Stueckelberg approach is equivalent, under certain conditions, to another approach that has been the topic of many articles: that of promoting constants of Nature to spacetime fields (see, e.g., \cite{Magueijo1,Magueijo2,Magueijo3,Magueijo4,Magueijo5,Magueijo6} and the related paper \cite{Magueijo7}, as well as the discussions in \cite{CG1,CG2,CG3,GT1,GT2,GT3,GT4}). This will be useful in the examples that follow.

Constants of Nature are quantities that enter as fixed parameters in the definition of a theory (i.e., in the action functional), and they are independent of space and time. If a parent gauge theory described by the action in Eq. \eref{action} has constants of Nature $\mathscr{P}_i$ ($i=1,\ldots,P$), one can trivially define a new theory with action functional
\eq{\label{eq:action-Nature}
S_{\rm new} &= \int_{\tau_0}^{\tau_1}\!\D\tau \Big\{\mathcal{A}_{\mu}[z,\mathscr{P}]\dot{z}^{\mu}+\mathscr{P}_i\dot{\mathscr{Q}}^i\\
&\ \ \ -H_0(\tau;z,\mathscr{P}]-\lambda^a\phi_a[z,\mathscr{P}]\Big\}\,,
}
which is identical to Eq. \eref{action} except for the fact that the constants of Nature $\mathscr{P}_i$ are promoted to fields (degrees of freedom), which are interpreted as the canonical momenta conjugate to cyclic degrees of freedom $\mathscr{Q}^i$. This then implies that the $\mathscr{P}$ fields do not change over time, and any initial condition will be frozen. If one assumes an initial value for the $\mathscr{P}$ fields that is constant across space (if $d>0$), then due to $\mathscr{P}$ conservation, the $\mathscr{P}$ solution to the field equations will be spatially constant at all times. On the other hand, an initial condition that changes across space would lead to ``constants'' of Nature with a static spatial variation. In any case, the phase space is extended to include the canonical pairs $(\mathscr{Q},\mathscr{P})$, and the new theory described by the action \eref{action-Nature} does not exhibit constants of Nature (theory-defining parameters) but rather only fields, some of which are conserved in time. Of course, we also assume that the constraints $\phi_a$ obey the regularity conditions with respect to the enlarged set of fields $(z,\mathscr{P},\mathscr{Q})$.

\subsubsection{Parametrization}
It is more convenient to express the action in Eq. \eref{action-Nature} as a completely constrained system, for which $H_0\equiv0$. In case $H_0$ is not identically zero, we can replace it with a constraint by a procedure called ``parametrization'' \cite{Kiefer:book}. 

First, if $d>0$, we now define $\mathtt{Z} = (z,\ldots,\partial^nz,\mathscr{P},\ldots,\partial^n\mathscr{P})$, and we assume that the unconstrained Hamiltonian can be written as (cf. Footnote \ref{foot:functionals})
\eq{\label{eq:fieldH}
H_0(\tau;z,\mathscr{P}] = \int\D^dx\ \mathscr{H}_0(\tau,\mathbf{x};{\mathtt{Z}(\tau,\mathbf{x})}) \ , 
}
and that the function $\mathscr{H}_0$ satisfies the first-class relation
\eq{\label{eq:field-first-class}
&\{\phi_a{(\mathtt{Z}(\tau,\mathbf{x}))},\mathscr{H}_0(\tau,\mathbf{x}';{\mathtt{Z}(\tau,\mathbf{x}')})\}\\
& \ \ = \sum_{b=1}^N\int\D^dy\,\mathscr{V}_a^b(\tau,\mathbf{x},\mathbf{x}',\mathbf{y};z,\mathscr{P}]\phi_b({\mathtt{Z}(\tau,\mathbf{y})}) \ ,
}
where we have made the summation over $b$ and the accompanying spatial integration explicit for clarity. With
\eq{
\mathcal{V}_a^b(\tau,\mathbf{x},\mathbf{y};z,\mathscr{P}]\Def \int\D^d x'\,\mathscr{V}_a^b(\tau,\mathbf{x},\mathbf{x}',\mathbf{y};z,\mathscr{P}] \ ,
}
then Eqs. \eref{fieldH} and \eref{field-first-class} yield
\eq{\label{eq:field-first-class2}
&\{\phi_a({\mathtt{Z}(\tau,\mathbf{x})}),H_0(\tau;z,\mathscr{P}]\}\\
& \ \ \ = \sum_{b=1}^N\int\D^dy\,\mathcal{V}_a^b(\tau,\mathbf{x},\mathbf{y};z,\mathscr{P}]\phi_b({\mathtt{Z}(\tau,\mathbf{y})}) \ ,
}
which coincides with the second of Eqs. \eref{first-class} after the summation and integration are again ommitted. To concatenate the notation, we formally define $\mathscr{H}_0$ to coincide with $H_0$ in the case $d=0$.

The procedure of parametrization then consists in introducing the new canonical pair $(T,p_T)$ and the global gauge constraint
\eq{\label{eq:par-phi0}
\phi_{0} = p_T+\mathscr{H}_0(T;\mathbf{x};{\mathtt{Z}}) \ ,
}
together with the global gauge-fixing condition
\eq{\label{eq:par-chi0}
\chi^{0} = T(\tau,\mathbf{x})-\tau \ , 
}
so as to reinterpret the action in Eq. \eref{action-Nature} as the restriction of\footnote{Evidently, the integral over $\mathbf{x}$ is absent if $d=0$.}
\eq{\label{eq:par-action-0}
S_{\rm new}^{\rm par} &= \int_{\tau_0}^{\tau_1}\!\D\tau\Biggl[\mathcal{A}_{\mu}\dot{z}^{\mu}+\mathscr{P}_i\dot{\mathscr{Q}}^i\\
&\ \ \ \ +\int\D^dx\,\left(p_T\dot{T}-\lambda^0\phi_0\right)-\lambda^a\phi_a\Biggr]
}
to the hypersurface defined by $\phi_{0}\approx0, \chi^{0}\approx0$. In this way, the field $T$ corresponds to a global clock that keeps track of the evolution.
 
If we include the $(T,p_{T})$ pair in the definition of the Poisson bracket in the usual way, it is straightforward to see that $\phi_{0}$ is a first-class constraint due to the original first-class algebra given by Eqs. \eref{first-class} and \eref{field-first-class}, and due to the fact that the $\phi_a$ constraints are assumed to have no explicit dependence on time (or $T$). We can then redefine the multipliers $\lambda_{\rm new} \Def (\lambda^{0},\lambda_{\rm old})$ and constraints $\phi_{\rm new} = (\phi_0,\phi_{\rm old})$, so as to obtain a new first-class algebra with no unconstrained Hamiltonian. As a matter of notation, we can also add $p_T$ to the list of $\mathcal{P}$ fields, and, correspondingly $T$ to the list of $\mathcal{Q}$ fields.\footnote{\label{foot:pT}This is just for convenience because, in general, $T$ will not be cyclic and $p_T$ will not be constant in time. If $\mathscr{H}_0$ or $H_0$ depend on $T$ explicitly, we have $\dot{p}_T(\tau,\mathbf{x})\approx-\int\D^dy \lambda^0(\tau,\mathbf{y})\delta\mathscr{H}_0|_{\mathbf{y}}/\delta T(\tau,\mathbf{x})$ if $d>0$ or $\dot{p}_T(\tau) \approx -\lambda^0(\tau)\partial H_0/\partial T$ if $d=0$.} In this way, the parametrized action in Eq. \eref{par-action-0} is equivalent to
\eq{\label{eq:par-action}
S_{\rm new}^{\rm par} &= \int_{\tau_0}^{\tau_1}\!\D\tau \left\{\mathcal{A}_{\mu}\dot{z}^{\mu}+\mathscr{P}_i\dot{\mathscr{Q}}^i-\lambda^a\phi_a[z,\mathscr{P}]\right\} ,
}
where the indices now run over the appropriate values for the enlarged set of variables together with implicit spatial integrations if $d>0$. We also assume that the enlarged set of constraints obeys the regularity conditions with respect to the $(z,\mathscr{P},\mathscr{Q})$ fields.

More generally, the procedure of parametrization of a theory consists in reinterpreting the action as the on-shell action of a larger gauge system by introducing new degrees of freedom and a system of first-class constraints [cf. Eq. \eref{par-phi0}] together with gauge-fixing conditions [cf. Eq. \eref{par-chi0}]. For convenience, we thus consider the action in Eq. \eref{par-action} (with $N$ first-class constraints $\phi_a$ that can be numbered $a=1,\ldots,N$ without loss of generality and $H_0\equiv0$) to be the starting point for a theory in which constants of Nature have been promoted to physical fields.\footnote{Eliminating the unconstrained Hamiltonian via parametrization is a matter of formalism. However, it is a physically relevant question whether a global clock field $T$ exists independently of the original $z$ degrees of freedom. If not, then the parametrization which brings the action \eref{action-Nature} into \eref{par-action-0} would seem to have no physical relevance. This question is also raised by the interpretation of the physical meaning of the Fock--Stueckelberg approach, as discussed in Sec. \ref{sec:conclusions}.}

\subsubsection{Solving the constraints in terms of constants of Nature}
The new action in Eqs. \eref{action-Nature} or \eref{par-action} offers the possibility of further extending the theory to include interactions of the $(\mathscr{Q},\mathscr{P})$ fields, thereby introducing notions of ``varying constants of Nature.''\footnote{It would be interesting to consider these interactions explicitly for some cosmological toy models. Of course, the idea of ``varying constants of Nature'' has been explored in several articles with different formalisms (see, e.g., \cite{varying1,varying2,varying3,varying4,varying5}).} However, even without introducing interactions, the promotion of the constants of Nature to degrees of freedom instead of mere theory-defining parameters has the advantage that one can now attempt to solve the constraints $\phi_a$ in terms of the $\mathscr{P}$ fields rather than the original $z$ fields. This may lead to a more straightforward solution, as it may be simpler to solve $\phi_a\approx0$ in terms of the $\mathscr{P}$ fields.

Let us assume that the number $P$ of constants of motion that have been promoted to fields is less than or equal to the number of constraints, $P\leq N$. Then, without loss of generality, we can also assume that the first $P$ constraints are solved in terms of the $\mathscr{P}$ degrees of freedom, which can then be written as
\eq{\label{eq:depar-constraint-0}
\mathscr{P}_a &\approx-\mathscr{H}_a(z,{\ldots,\partial^nz,}\mathscr{P}_{i>a}{,\ldots,\partial^n\mathscr{P}_{i>a}}) \ ,\\
 &\ (a=1,\ldots,P) \ ,
}
in a region of phase space. One can then use these solutions recursively to express $\mathscr{H}_a$ only in terms of the $z$ fields and their derivatives. In the region of phase space where this holds, the constraint hypersurface can be defined by the new set of constraints\footnote{\label{foot:par-depar}The parametrization constraint in Eq. \eref{par-phi0} is not necessarily a particular case of Eq. \eref{depar-constraint} because $p_T$ need not be constant in time, i.e., a constant of Nature (cf. Footnote \ref{foot:pT}). Furthermore, notice that while the constraints that come from the parametrization procedure, such as the one in Eq. \eref{par-phi0}, may be globally valid in phase space, the solutions in Eqs. \eref{depar-constraint} may hold only in a certain region. Thus, in general, a parametrized theory with constraints of the form given in Eq. \eref{depar-constraint} may not be globally equivalent to gauge theory with constraints $\phi_a$, with the equivalence holding locally in certain cases. We also disregard a possible explicit spatial dependence of $\mathscr{H}_a$ in Eqs. \eref{depar-constraint-0} and \eref{depar-constraint}.}
\eq{\label{eq:depar-constraint}
\bar{\phi}_a &\Def \mathscr{P}_a+\mathscr{H}_a(z,{\ldots,\partial^nz}) \approx 0 \ , \ (a=1,\ldots,P) \ ,\\
\bar{\phi}_b&\Def \phi_b \approx 0 \ , \ (b=P+1,\ldots,N) \ .
} 

\subsubsection{\label{sec:constraint-redef}General constraint redefinitions}
What can we conclude about the constraint transformation in Eqs. \eref{depar-constraint}? If we assume that this transformation is invertible in a region of phase space in a neighborhood of the constraint hypersurface, then it is a particular case of a general transformation $\phi_a\mapsto \bar{\phi}_a(\phi)$ of constraints that is well defined and invertible in a neighborhood of the constraint hypersurface and satisfies $\bar{\phi}_a = 0$ if, and only if, $\phi_a = 0$ (for every $a=1,\ldots,N$). Any such transformation must satisfy $\bar{\phi}_a = \tensor{M}{_a^b}\phi_b$ due to Eq. \eref{KT-0}, with $\tensor{M}{_a^b}$ being the coefficients of a matrix that must be non-divergent and invertible at least in a neighborhood of the constraint hypersurface. The new constraints $\bar{\phi}_a$ are thus independent, and they determine the same hypersurface as the original $\phi_a$ functions, but they will generally have different structure functions.

In the general case, the constraint transformation induces a transformation of the multipliers, $\bar{\lambda}^a = \lambda^b\tensor{\left(M^{-1}\right)}{_b^a}$, so that we may write the Hamiltonian as $H = \lambda^a\phi_a = \bar{\lambda}^a\bar{\phi}_a$. This redefinition of $\phi_a$ and $\lambda^a$ is not a gauge transformation because the functional form of the constraints will generally change, and the transformation of $\lambda$ will generally not involve a time derivative of an arbitrary function in the form of Eq. \eref{delta-lambda}.

As a particular case, we assume that the new constraints in Eqs. \eref{depar-constraint} are functionally independent and that they obey the regularity conditions. To verify the relation $\bar{\phi}_a = \tensor{M}{_a^b}\phi_b$, one may expand $\phi_a$ ($a=1,\ldots,P$) in a Taylor series in powers of $\mathscr{P}$ to find, in particular,
\eq{
\phi_a = \phi_a|_{\bar{\phi} = 0}+\del{\phi_a}{\mathscr{P}_b}\bar{\phi}_b+\frac12\del{^2\phi_a}{\mathscr{P}_b \mathscr{P}_c}\bar{\phi}_b\bar{\phi}_c+\mathcal{O}(\bar{\phi}^3) \,,
}
where $a,b = 1,\ldots,P$, $\bar{\phi}_a$ are given by Eq. \eref{depar-constraint}, and we suppressed the arguments of $\phi_a$ for brevity. Then, since $\phi_a|_{\bar{\phi}}=0$, we have
\eq{
\phi_a &= \bar{\phi}_b\left[\del{\phi_a}{\mathscr{P}_b}+\frac12\del{^2\phi_a}{\mathscr{P}_b \mathscr{P}_c}\bar{\phi}_c+\mathcal{O}(\bar{\phi}^2)\right]\\
& =: \tensor{\left(M^{-1}\right)}{_a^b}\bar{\phi}_b \ , \ (a,b=1,\ldots,P) \ .
}
As we assume that $\phi_a\approx0$ ($a=1,\ldots,P$) if, and only if, the constraints in Eq. \eref{depar-constraint-0} hold, then the matrix $\tensor{\left(M^{-1}\right)}{_a^b}$ must be invertible on the constraint hypersurface. In a neighborhood of this hypersurface, for $a,b,c=1,\ldots,P$, we can define $\frac12\partial^2\phi_a/\partial\mathscr{P}_b\partial\mathscr{P}_c$ to be the element of row $a$ and column $b$ of a matrix $\mathtt{H}^c$ for each value of $c$, so that $\det M^{-1} = \det M^{-1}|_{\bar{\phi}=0}(1+\bar{\phi}_c\mathrm{Tr}(M\mathtt{H}^c)|_{\bar{\phi}=0}+\mathcal{O}(\bar{\phi}^2))$, which is nonvanishing for $\bar{\phi}$ with sufficiently small but nonzero absolute values. We can then trivially extend $\tensor{(M^{-1})}{_a^b}$ to $a,b=1,\ldots,N$ because the constraints $\phi_a$ with $a=P+1,\ldots,N$ are not changed in this particular case. The theory can thus be defined from the action
\eq{\label{eq:par-action-depar}
S_{\rm new}^{\rm depar} &= \int_{\tau_0}^{\tau_1}\!\D\tau \left\{\mathcal{A}_{\mu}\dot{z}^{\mu}+\mathscr{P}_i\dot{\mathscr{Q}}^i-\bar{\lambda}^a\bar{\phi}_a[z,\mathscr{P}]\right\} ,\!
}
with the new constraints and multipliers, which is locally equivalent to Eq. \eref{par-action}. In Sec. \ref{sec:equiv-constant-FS}, we will use this new theory as the parent gauge theory in the Fock--Stueckelberg approach.

\subsubsection{Deparametrization}
A transformation of constraints that involves the solution of the original constraints in terms of a momentum variable is called ``deparametrization.''\footnote{This is conceptually the inverse procedure of parametrization. However, the two procedures may lead to theories that are not globally equivalent. See Footnote \ref{foot:par-depar}.} The deparametrization in Eqs. \eref{depar-constraint} motivates a gauge fixing given by the conditions
\eq{\label{eq:depar-gf}
\bar{\chi}^a&\Def \mathscr{Q}^a-f^a \approx 0 \ ,
}
where $f^a$ are functions solely of the spacetime point. These gauge conditions fix the first $P$ multipliers to be [cf. Eq. \eref{fix-mult}]:
\eq{\label{eq:depar-fix-mult}
\bar{\lambda}^a \approx \dot{f}^a \equiv \bar{\Lambda}^a \ .
}
If we replace Eqs. \eref{depar-constraint-0} and \eref{depar-gf} in the action in Eq. \eref{par-action-depar}, we obtain
\eq{\label{eq:depar-action}
S_{\rm new}^{\rm par} &= \int_{\tau_0}^{\tau_1}\!\D\tau\left[\mathcal{A}_{\mu}\dot{z}^{\mu}-H_{\rm red}\right] \,,\\
H_{\rm red} &\Def \sum_{a=1}^P\mathscr{H}_a\dot{f}^a+\sum_{b = P+1}^N\lambda^b\phi_b \,.
}
This corresponds to the action functional of a system with Hamiltonian $H_{\rm red}$, understood as a function solely of the spacetime point and $(z,\ldots,\partial^{n}z)$.\footnote{Recall that all constants of Nature that were promoted to fields have been replaced by the corresponding $-\mathscr{H}_a$, and their conjugate cyclic variables were replaced by $f^a$. Of course, $H_{\rm red}$ may still depend on other constant parameters that are not written explicitly.} Thus, the solution of the constraints and the gauge fixing in Eqs. \eref{depar-constraint-0} and \eref{depar-gf} effectively yield a system with a reduced set of degrees of freedom [the $z$ fields instead of the set $(z,\mathscr{Q},\mathscr{P})$]. For this reason, the Hamiltonian $H_{\rm red}$ is sometimes referred to as the `reduced Hamiltonian' \cite{Kiefer:book}. Likewise, the gauge freedom related to the constraints $\phi_a$ ($a=1,\ldots,P$) and its associated ``gauge parameters'' that change with gauge transformations have been removed: that is why one says that the system has been (partially) ``deparametrized,'' and Eqs. \eref{depar-constraint} are the deparametrized form of the constraints $\phi_a$ ($a=1,\ldots,P$). If we can set $f^{a_0} = \tau$, then the field $\mathscr{Q}^{a_0}$ plays the role of a physical clock. We will see in Sec. \ref{sec:equiv-constant-FS} how this is related to the Fock--Stueckelberg approach, and we discuss such a deparametrization in the context of general relativity in Secs. \ref{sec:GRconstants} and \ref{sec:reappear}.

\subsubsection{Physical meaning of promoting constants of Nature to spacetime fields}
If the quantities $\mathscr{P}$ are constants of Nature (i.e., fixed theory-defining parameters), then the possibility to write the constraints in the form given in Eq. \eref{depar-constraint} makes it clear that the $z$ fields cannot take all possible initial values but only those for which the quantities $\mathscr{H}_a$ take the fixed (theory-defining) values $-\mathscr{P}_a$. This is the traditional interpretation of first-class constraints in a gauge theory (cf. Sec. \ref{sec:gauge-theory}): they restrict the initial-value problem of the theory's degrees of freedom (the $z$ fields).

With the promotion of $\mathscr{P}$ to degrees of freedom on a par with the $z$ fields, the problem is turned upside down: rather than a restriction on the initial $z$ values, the constraints in Eq. \eref{depar-constraint} should be viewed as restrictions on the otherwise unspecified values of $\mathscr{P}$. In this way, the $z$ fields can now take all possible initial values, as different choices (different solutions) simply amount to assigning different $\mathscr{P}$ values. In this sense, the original theory's constraints on the $\mathscr{H}_a$ (seen as functions of $z$) are relaxed. This is exactly the same result of the Fock--Stueckelberg approach, to which we now turn.

\subsubsection{\label{sec:equiv-constant-FS}Equivalence with the Fock--Stueckelberg theory}
Let us take the action in Eq. \eref{par-action-depar} as the definition of the parent gauge theory, and Eq. \eref{depar-fix-mult} is taken as the definition of the noncanonical gauge condition (cf. Sec. \ref{sec:noncanon}). Then, the Fock--Stueckelberg action with $H_0\equiv0$ given in Eq. \eref{FS-action-1} becomes
\eq{\label{eq:FS-action-constants}
\!\tilde{S}_{\rm FS} \Def \!\int_{\tau_0}^{\tau_1}\!\!\!\D\tau\! \left[\mathcal{A}_{\mu}\dot{z}^{\mu}+\mathscr{P}_i\dot{\mathscr{Q}}^i-\sum_{a=1}^P\dot{f}^{a}\bar{\phi}_{a}-\!\!\!\sum_{b=P+1}^{N}\!\!\lambda^{b}\phi_{b}\right] .
}
The $z$-field equations then read [cf. Eqs. \eref{FS-eom}]
\eq{\label{eq:z-eom-constants}
\dot{z}^{\mu} &= \left\{z^{\mu},\sum_{a=1}^P\dot{f}^{a}\mathscr{H}_a+\sum_{b=P+1}^{N}\!\!\lambda^{b}\phi_{b}\right\}\\
&=\{z^{\mu},H_{\rm red}\}
}
because $f^a$ are functions solely of the spacetime point, and $z$ Poisson-commutes with $\mathscr{P}$ by definition. The equations \eref{z-eom-constants} are exactly those derived from the action in Eq. \eref{depar-action}. In this way, as far as the $z$ fields are concerned, both the Fock--Stueckelberg and constants of Nature approaches are equivalent. This is of course expected because the deparametrization via constants of Nature induced a choice of gauge [cf. Eqs. \eref{depar-gf} and \eref{depar-fix-mult}], which can then be used to seed the Fock--Stueckelberg approach. However, it is important to emphasize that this equivalence is established starting from the action in Eq. \eref{par-action-depar}, in which the constraints are already in deparametrized form. As this form holds only locally in general, the Fock--Stueckelberg theory defined from a more general parent theory, with action given in Eq. \eref{par-action} for example, may generally exhibit global differences. Whereas the physical meaning of the constants of Nature approach becomes most clear with deparametrization,\footnote{In the constants of Nature approach, the effects of the relaxation of constraints, as well as the resulting physical equations of motion derived from the action in Eq. \eref{depar-action}, are most clearly seen if the deparametrization can be explicitly performed.} the effects of the Fock--Stueckelberg approach on the relaxation of constraints are evident without deparametrization, as they only require a well-defined noncanonical gauge condition.

\subsection{\label{sec:subalgebra}First-class subalgebras}
Let us consider a parent gauge theory with the following property: we can divide the constraints into $\phi = (u,K)$ such that $u_{a} = \phi_a$ ($a = 1,\ldots, N_u$) and $K_{a} = \phi_{N_u+a}$ ($a = 1,\ldots, N-N_u$), and we assume that the structure functions have the property $\mathcal{V}^c_b = 0, \mathcal{F}_{ab}^c = 0$ if $c = 1,\ldots,N_u$. If this is not the case, it may be possible to bring the structure functions into this form by a redefinition of the constraints (cf. Sec. \ref{sec:constraint-redef}),\footnote{In local regions of field space, it may be possible to choose $\tensor{M}{_a^b}$ in $\bar{\phi}_a = \tensor{M}{_a^b}\phi_b$ (cf. Sec. \ref{sec:constraint-redef}) so that the new structure functions vanish, and the new constraints locally satisfy an Abelian algebra $\{\bar{\phi}_a,\bar{\phi}_b\} = 0$. In general, however, it is not possible to achieve this `Abelianization' in a global manner \cite{HT:book}.} and the $u$ constraints Poisson-commute if the $K$ constraints are satisfied:\footnote{Indices contracted with the index of $\phi_a$ are summed from $1$ to $N$. However, indices contracted with the index of $u_a$ are to be summed from $1$ to $N_u$ only, whereas indices contracted with the index of $K_a$ are to be summed from $1$ to $N-N_u$.}
\eq{\label{eq:subalgebra}
\{\phi_a,\phi_b\} &= \mathcal{F}_{ab}^{N_u+c}K_c \ ,\\
\{\phi_a,H_0\} &= \mathcal{V}_a^{N_u+b}K_b \ .
}
The $K$ functions lead to a first-class subalgebra.\footnote{The structure functions may in principle depend on all degrees of freedom and may have some functional dependence also on the $u$ functions.} In particular, the relations \eref{subalgebra} also lead to the field equations
\eq{\label{eq:subalgebra-evol}
\dot{\phi}_a = \mathcal{V}_a^{N_u+b}K_b+\lambda^b\mathcal{F}_{ab}^{N_u+c}K_c+\{\phi_a,\lambda^b\}\phi_b \ .
}
If we can locally define phase-space functions $\xi^a$ ($a=1,\ldots,N_u$), which satisfy $\{\xi^b,u_a\} \approx \delta^b_a$ for $a,b=1,\ldots,N_u$ and $\{\xi^b,H_0\}\approx\{\xi^b,K_a\} \approx 0$, then we can consider the partial gauge fixing:
\eq{
\chi^a\Def \xi^a-f^a(\tau,\x)\ ,\ (a=1,\ldots,N_u) \ ,
}
which is preserved in time if we have
\eq{\label{eq:gauge-stability-subalgebra}
0&\stackrel{\chi^a=0}{=}\dot{\chi}^a = \del{\chi^a}{\tau}+\{\chi^a,H_0+\lambda^b u_b+\lambda^{N_u+b}K_b\}\\
&=\del{\chi^a}{\tau}+\lambda^a+\{\chi^a,\lambda^b\}u_b+\{\chi^a,\lambda^{N_u+b}\}K_b
}
on the hypersurface defined by $\chi^a\approx0$. If we also enforce $K_b\approx0$, then Eq. \eref{gauge-stability-subalgebra} can be solved for $\lambda^a$ ($a=1,\ldots,N_u$) similarly to Eq. \eref{fix-mult} by fixing the first $N_u$ multipliers to be:
\eq{\label{eq:subalgebra-fix-mult}
\lambda^a = \dot{f}^a(\tau,\x) \ ,\ (a=1,\ldots,N_u)\ .
}
Notice that we only used $\chi^a,K_b\approx0$ in this result, and we did not need to enforce $u_a \approx 0$. Using this partial gauge condition in Eq. \eref{subalgebra-evol}, we find that the $u_a$ constraints evolve as
\eq{\label{eq:subalgebra-evol-u}
\dot{u}_a = \mathcal{V}_a^{N_u+b}K_b+\lambda^b\mathcal{F}_{ab}^{N_u+c}K_c+\{u_a,\lambda^{N_u+b}\}K_b \ ,
}
and they are conserved if $K_a\approx0$ is enforced, regardless of whether $u_a \approx 0$ is imposed.

The corresponding Fock--Stueckelberg theory with the noncanonical gauge conditions in Eq. \eref{subalgebra-fix-mult} leads to the constraints $\lambda^a\approx\dot{f}^a(\tau,\x), u_a+w_a\approx0, K_b\approx0$ [cf. Eqs. \eref{FS-constraints}]. Due to the conservation of $u_a$ on the hypersurface defined by $K_b\approx0$, the effect of the Fock--Stueckelberg constraint $u_a+w_a\approx0$ is simply to assign a {constant-in-time} value to the $w_a$ multiplier, and the constraints $u_a$ are relaxed: instead of being forced to take the value $0$, they can now take any {constant-in-time} value. This is useful because many models of interest, such as (bosonic) relativistic particles (which have $d=0,H_0\equiv0$ and an Abelian algebra of constraints), can be seen as particular cases of this example.

\subsection{\label{sec:mechsimple}A simple mechanical example}
Now consider a mechanical $(d=0)$ theory in which the $z$ degrees of freedom correspond to canonical pairs $(x,p_x,y,p_y)$ with symplectic potential $\mathcal{A} = (p_x/2,-x/2,p_y/2,-y/2)$ and an action that can be written as (up to a boundary term)
\eq{
S = \int_{\tau_0}^{\tau_1}\!\D\tau \left(p_x\dot{x}+p_y\dot{y}-\frac{p_x^2}{2}-\frac{p_y^2}{2}-\lambda p_x\right) .
}
One can verify that the single constraint is of first class and Abelian, and we thus have a particular case of the example discussed in Sec. \ref{sec:subalgebra}. The solution to the equations of motion is readily found to be
\eq{\label{eq:simplesol}
x(\tau) &= x_0+\int^\tau\!\D\tau'\lambda(\tau') \ , \ p_x(\tau) = 0 \ , \\
y(\tau) &= y_0+k\tau \ , \ p_y(\tau) = k\ ,
}
with $x_0, y_0$ and $k$ being constants. As $x$ depends on the arbitrary function $\lambda(\tau)$, it is ``pure gauge;'' i.e., it only encodes information concerning the generalized reference frame, and it does not have any gauge invariant content. Indeed, its gauge transformation is $\delta x = \varepsilon(\tau)$, which can be integrated to $x(\tau;s) = x(\tau)+s\varepsilon(\tau)$, where $s$ is the parameter along the `gauge orbit', which is the integral curve of the gauge generator. Thus, the gauge condition $\chi = x(\tau) \approx 0$ can be used to eliminate $x(\tau)$, as it is both accessible and complete because, with $\varepsilon\Def -x(\tau)$ being independent of $s$, the initial arbitrary generalized reference frame can be mapped into the frame with vanishing $x(\tau)$ at $s=1$,\footnote{The value of $s$ at which the gauge condition is reached of course depends on $\varepsilon$. Here, it is set to $1$ for convenience.} and only the trivial gauge transformation (with $\varepsilon\equiv0$) preserves the gauge condition. In this way, the only degrees of freedom of this model are $y_0$ and $k$.

The canonical gauge $x(\tau) \approx 0$ implies the noncanonical gauge $\lambda(\tau)\approx0$.\footnote{This follows from Eq. \eref{simplesol} or from the equation of motion $\dot{x}=p_x+\lambda\approx\lambda$. Notice also that Eq. \eref{delta-lambda} leads to the gauge transformation of the multiplier $\delta\lambda=\dot{\varepsilon}$ or, in integrated form, $\lambda(\tau;s) = \lambda(\tau)+s\dot{\varepsilon}(\tau)$. Starting from an arbitrary generalized reference frame, the gauge transformation with $\varepsilon\Def-x(\tau)$ (which is independent of $s$) leads to $\lambda(\tau;s=1) \approx 0$, as it should. The converse, however, is not true: the noncanonical gauge $\lambda\approx0$ does not imply the canonical gauge $x(\tau)\approx0$ because $\lambda\approx0$ is not complete. It is preserved by the residual gauge transformations with $\varepsilon=\,$const., and thus $\lambda\approx0$ only implies that $x(\tau)\approx$ const..} In this case, the Fock--Stueckelberg action \eref{FS-action-1} becomes
\eq{
\!\tilde{S}_{\rm FS} = \int_{\tau_0}^{\tau_1}\!\D\tau \left[p_x\dot{x}+p_y\dot{y}-\frac{(p_x^2+p_y^2)}{2}\right] ,
}
which readily leads to the solutions
\eq{
x(\tau) &= x_0+k_x\tau \ , \ p_x(\tau) = k_x \ , \\
y(\tau) &= y_0+k\tau \ , \ p_y(\tau) = k\ .
}
This theory clearly has more degrees of freedom (and less gauge invariance), as the $x_0, y_0, k_x, k$ constants are freely specifiable. The parent gauge theory is only recovered for the set of trajectories which obey $k_x = 0$, and thus only a particular class of solutions of the Fock--Stueckelberg theory corresponds to the solutions of the parent theory in the noncanonical gauge $\lambda\approx0$.

\subsection{\label{sec:mech}To gauge or not to gauge time?}
Mechanical theories ($d = 0$) without local symmetries can be defined from the action [cf. \eref{action}]
\eq{\label{eq:mechaction1}
S_{\rm mech} \Def \int_{\tau_0}^{\tau_1}\D\tau\ \left[\mathcal{A}_{\mu}\dot{z}^{\mu}-H(z)\right] \ ,
}
which forms the starting point of the usual Hamilton variational principle. The particle's degrees of freedom, given by the $z^{\mu}$ ($\mu = 1,\ldots,Z$) variables, are scalar fields on its worldline, and $\tau$ is an `absolute' or `external' time parameter. Since the Hamiltonian is assumed to have no explicit time dependence, it is conserved, and time translations form a global (rigid) symmetry. They can be `gauged' (i.e., turned into a local symmetry) if we introduce a worldline scalar density $\lambda(\tau)$ as an independent auxiliary field, $H(z)\to \lambda H(z)$, so as to make both terms in the Lagrangian (the symplectic potential term and the Hamiltonian) transform as worldline scalar densities. The new action
\eq{\label{eq:mechaction2}
\tilde{S}_{\rm mech} \Def \int_{\tau_0}^{\tau_1}\D\tau\ \left[\mathcal{A}_{\mu}\dot{z}^{\mu}-\lambda H(z)\right]
}
is that of a canonical gauge system, since the equation of motion of $\lambda$ is the Hamiltonian constraint, $H \approx 0$, which is clearly of first class and Abelian. The result is a gauge theory of time translations: the parameter $\tau$ is no longer absolute, as it may be redefined [$\tau\to\tau+\varepsilon(\tau)$] without changing the form of the equations of motion, which are said to be time-reparametrization invariant. This is a particular case of the example considered in Sec. \ref{sec:subalgebra}.

The arbitrariness of $\tau$ is of course inherited from the arbitrariness of $\lambda$, and the fixation of this `generalized reference frame' is a gauge choice. In particular, with canonical gauge conditions, one can define time internally by fixing $\tau$ to be a function of $z$, the particle's degrees of freedom \cite{Hanson,HT:book,Chataig:Thesis}. This, together with the Hamiltonian constraint $H\approx0$, means that $2$ out of the $Z$ degrees of freedom of the particle are not independent of the others. In this way, by gauging the time translations, we exclude independent degrees of freedom.

We can restore the eliminated degrees of freedom with the Fock--Stueckelberg mechanism. With the noncanonical gauge condition $\chi = \lambda-\Lambda$ [cf. Eqs. \eref{FS-gf-main}], the Fock--Stueckelberg action \eref{FS-action} reads
\eq{\label{eq:mechFS1}
S_{\rm FS} \Def \int_{\tau_0}^{\tau_1}\D\tau\ \left[\mathcal{A}_{\mu}\dot{z}^{\mu}-\lambda H(z)-w(\lambda-\Lambda)\right] \ ,
}
and it leads to the constraints $H +w\approx 0$, $\lambda\approx\Lambda$, which are satisfied at all times. Since $w$ is an extra degree of freedom, the Hamiltonian constraint is relaxed, and the equations of motion are equivalent to those derived from the action given in Eq. \eref{FS-action-1}, which reads
\eq{\label{eq:mechFS2}
\tilde{S}_{\rm FS} \Def \int_{\tau_0}^{\tau_1}\D\tau\ \left[\mathcal{A}_{\mu}\dot{z}^{\mu}-\Lambda H(z)\right] \ .
}
This coincides with the original Hamilton-principle action given in Eq. \eref{mechaction1} if $\Lambda$ is chosen to be a constant equal to $1$. Due to the relaxation of the Hamiltonian constraint and the loss of the symmetry under local time reparametrizations, the $2$ eliminated degrees of freedom are restored. Moreover, to return to the parent theory, one can promote $\Lambda$ to an arbitrary multiplier in Eq. \eref{mechFS2}, or set $w\equiv0$ in Eq. \eref{mechFS1}.

\subsubsection{\label{sec:relpart}The relativistic particle: from Jacobi to Hamilton and back}
A particular case of this mechanical theory can be used to describe a relativistic particle. If $z = (q,p)$ are canonical pairs with symplectic potential $\mathcal{A} = (p/2,-q/2)$ and Hamiltonian\footnote{The Latin indices for the relativistic particle variables are not to be confused with the multiplier indices used in the previous sections. In particular, we have $q^0 = ct$, with $c$ being the speed of light and $t$ being the time coordinate of the spacetime background on which the particle moves. The background spacetime metric $g_{ab}$ is taken with a mostly plus signature.}
\eq{\label{eq:relpartH}
H(q,p) &= \frac12 g^{ab}(q)\left[p_a+f_a(q)\right]\left[p_b+f_b(q)\right]\\
&\ \ \ +m(V(q)-\mathcal{E})\ ,
}
then we have the time-reparametrization invariant equation of motion $\dot{q}^a = \lambda g^{ab}\left(p_b+f_b\right)$, and so the phase-space action given in Eq. \eref{mechaction2} (which is defined as a functional of paths on the cotangent bundle of configuration space) is equivalent to the configuration-space tangent-bundle action
\eq{\label{eq:mechaction3}
S_{\rm particle} =  \int_{\tau_0}^{\tau_1}\D\tau \left[\frac{g_{ab}\dot{q}^a\dot{q}^b}{2\lambda} -f_{a}\dot{q}^a-\lambda m (V-\mathcal{E})\right]
}
after a Legendre transform. In these variables, the equation of motion of $\lambda$ (the constraint) becomes
\eq{\label{eq:LagrangianConstraint}
-\frac{g_{ab}\dot{q}^a\dot{q}^b}{2\lambda^2}-mV+m\mathcal{E} = 0 \ .
}
By solving it for $\lambda$ and replacing the solution in Eq. \eref{mechaction3}, we obtain 
\eq{\label{eq:mechaction4}
S_{\rm Jacobi} \!=\!  \int_{\tau_0}^{\tau_1}\D\tau \left[-\sigma\sqrt{2m(\mathcal{E}-V)g_{ab}\dot{q}^a\dot{q}^b} - f_{a}\dot{q}^a\right] ,
}
where $\sigma=\pm1$ is the sign of $\lambda$, which plays the role of the `einbein' on the particle's worldline. If the function $f_a(q)\equiv0$, and with $\mathcal{E}$ being a constant, this is precisely the usual Jacobi action (up to an overall sign) that can be used to define time-reparametrization invariant trajectories in classical mechanics via its extremization (the Jacobi action principle)\cite{Lanczos}.\footnote{Starting from the Arnowitt--Deser--Misner (ADM) formulation of general relativity [discussed in \cite{Kiefer:book,ADM} (and references therein) and briefly in Sec. \ref{sec:gug}], a similar procedure leads to a square-root action for general relativity, known as the Baierlein--Sharp--Wheeler action \cite{Kiefer:book,Baierlein}. Analogously, starting from the Polyakov action for a relativistic string, one can obtain a square--root action known as the Nambu--Goto action \cite{strings1,strings2,strings3,strings4}.} Furthermore, if $\sigma=1$, $f_a = -e A_a$, $V(q)\equiv0$ and $\mathcal{E}\equiv -mc^2/2$, the action in Eq. \eref{mechaction4} reduces to the usual action for a relativistic particle with charge $e$ coupled to an external electromagnetic field. In this case, the corresponding Hamilton--Jacobi equation reads
\eq{\label{eq:HJ-Dirac-gauge}
g^{ab}\left[\del{S}{q^a}-e A_a\right]\left[\del{S}{q^b}-e A_b\right]+m^2c^2 =0\ ,
}
which will be useful in Sec. \ref{sec:DiracEM}. In any case, the time-reparametrization invariance is manifest in the action given in Eq. \eref{mechaction4}. 

With the Fock--Stueckelberg mechanism [cf. Eq. \eref{mechFS1}], we are led from the parent theory defined by Eq. \eref{mechaction3} [which is equivalent to Eq. \eref{mechaction4}] to a theory defined by
\eq{\label{eq:mechaction5}
S_{\rm Hamilton} =  \int_{\tau_0}^{\tau_1}\D\tau \left(\frac{m}{2}g_{ab}\dot{q}^a\dot{q}^b -f_{a}\dot{q}^a-V+\mathcal{E}\right)\,,
}
where we have taken the noncanonical gauge $\lambda = 1/m$ for simplicity.\footnote{For the ordinary relativistic free particle, for which $\sigma=1$, $V \equiv 0$, $\mathcal{E} = -mc^2/2$ and $f_a\equiv0$, we find from Eq. \eref{LagrangianConstraint} that the gauge $\lambda=1/m$ corresponds to fixing $\tau$ to be the particle's proper time (on a solution to the equations of motion), while the constraint in Eq. \eref{LagrangianConstraint} enforces that the particle travels on a timelike trajectory. For a massless particle ($m=0$), we cannot define the Jacobi action [cf. \eref{mechaction4}]. Nevertheless, the action in Eq. \eref{mechaction3} is well defined, and we can transition to an action akin to the one in Eq. \eref{mechaction5} with the noncanonical gauge $\lambda = \Lambda =\,$const., where the constant $\Lambda$ has units of inverse mass. This corresponds to fixing $\tau$ to be an affine parameter for the null trajectory of the massless particle.} The action in Eq. \eref{mechaction5} is equivalent, via a Legendre transform, to an action of the form of Eq. \eref{mechFS2} with $\Lambda\equiv1/m$, and it is the action used in the ordinary Hamilton variational principle for unconstrained mechanics. Thus, we see that the definition of the Fock--Stueckelberg theory for a relativistic particle corresponds to a transition from the variational principle of Jacobi to that of Hamilton, whereupon time translations are no longer gauged.\footnote{As discussed above, from the point of view of the parent theory, the Fock--Stueckelberg theory is obtained from an ``erroneous'' gauge fixing that restores degrees of freedom.}

The Hamiltonian derived from Eq. \eref{mechaction5} is identical to the one in Eq. \eref{relpartH}, but it is not constrained to vanish. Rather, due to the global time-translation invariance of Eq. \eref{mechaction5}, it is simply conserved $H(q(\tau),p(\tau)) = H(q(0),p(0))$. To transition back from the Hamilton action principle defined by Eq. \eref{mechaction5} to the Jacobi principle derived from Eq. \eref{mechaction4}, it suffices to restrict oneself to initial values for which the Hamiltonian vanishes,
\eq{\label{eq:back-to-Jacobi-0}
H(q(\tau),p(\tau)) = H(q(0),p(0)) \to 0 \ ,
}
which is equivalent to
\eq{\label{eq:back-to-Jacobi}
\frac{m}{2} g_{ab}\dot{q}^a\dot{q}^b+V-\mathcal{E} = \text{const.}\to 0 \ ,
}
as this recovers the Hamiltonian constraint [cf. Eq. \eref{LagrangianConstraint}] as a particular initial-value condition on the Fock--Stueckelberg theory.\footnote{This is equivalent to setting $w\equiv0$ in Eq. \eref{mechFS1}, as $w$ is constrained to be the value of $H$ on shell.} This condition is also responsible for the ensuing time-reparametrization invariance of the trajectories that exhibit such initial values.

It is also worthwhile to notice that the Fock--Stueckelberg trajectories of a relativistic particle (e.g., with $V\neq0$) need not be timelike or null throughout configuration space. Indeed, the transition from the parent to the Fock--Stueckelberg theory allows more solutions to the equations of motion: whereas the parent theory constrains initial-values, these are free in the Fock--Stueckelberg theory. This allowed Stueckelberg to consider trajectories going forwards and backwards in Minkowski time, which was useful in the description of pair creation and annihilation \cite{Stueck,Stueck1,Stueck2}.

\subsubsection{\label{sec:time-conjugate-constant}Time as the canonical conjugate to constants of Nature and the reappearance of time in the quantum theory}
For the Fock--Stueckelberg trajectories that do not satisfy the initial-value Hamiltonian constraint, we find from Eq. \eref{back-to-Jacobi} that the quantity
\eq{\label{eq:relpart-FS-conservation}
\frac{m}{2} g_{ab}\dot{q}^a\dot{q}^b+V = \text{const.}
}
is conserved because $\mathcal{E}$ is assumed to be a constant parameter (e.g., $\mathcal{E} = -mc^2/2$). The conserved value of Eq. \eref{relpart-FS-conservation} is, however, free, as it depends on a choice of initial conditions.

On the other hand, for the trajectories that respect the initial-value Hamiltonian constraint, we obtain from Eqs. \eref{back-to-Jacobi-0} and \eref{back-to-Jacobi} the equivalent conditions
\eq{
&\frac{m}{2} g_{ab}\dot{q}^a\dot{q}^b+V = \mathcal{E} \ ,\\
& \frac{g^{ab}}{2m} \left(p_a+f_a\right)\left(p_b+f_b\right)+V = \mathcal{E} \ .
}
The parameter $\mathcal{E}$ can be regarded as a ``constant of Nature'' of this toy theory. If we regard it as a new degree of freedom that is conserved, we can trivially enlarge the set of variables from $(q,p)$ to $(q,p;\eta,\mathcal{E})$, where $(\eta,\mathcal{E})$ is a new canonical pair (cf. Sec. \ref{sec:constants}). We can add $-\mathcal{E}\dot{\eta}$ to the action, so that the new time-reparametrization invariant action can be written as
\eq{
S_{\mathcal{E}} =  \int_{\tau_0}^{\tau_1}\D\tau \left[-\mathcal{E}\dot{\eta}+p_a\dot{q}^a-\lambda H(q,p;\mathcal{E})\right] \ .
}
As $\eta$ evolves linearly in the ``proper time'' $m\int^{\tau}\!\D\tau'\!\lambda(\tau')$,\footnote{For the free massive relativistic particle, with $f_a \equiv V \equiv0$ and $\mathcal{E}\equiv -mc^2/2$, one can verify that this quantity indeed coincides with the usual definition of the particle's proper time (up to an overall sign associated with the sign of the chosen $\lambda$; e.g., $\lambda=1/m$ has a positive sign, as $m>0$).} it can be taken to be a physical clock or a physical definition of time. In this sense, time can be seen as a physical variable canonically conjugate to a constant of Nature. For the time-reparametrization invariant parent theory, we can thus consider the canonical gauge condition $\chi = \eta(\tau)-\tau$, which fixes the particle's worldline time parameter to be $\eta$. Concomitantly, we can solve the Hamiltonian initial-value constraint not in terms of the original pairs $(q,p)$ but rather in terms of $\mathcal{E}$ -- the solution is trivial. The on-shell, gauge-fixed action becomes
\eq{
S_{\mathcal{E}} =  \int_{\tau_0}^{\tau_1}\!\!\!\D\tau \left\{p_a\dot{q}^a-\left[\frac{g^{ab}}{2m} \left(p_a+f_a\right)\left(p_b+f_b\right)+V\right]\right\} .
}
This is an unconstrained action of the type of the usual Hamilton principle. Its global time-translation invariance leads exactly to the conservation law expressed in Eq. \eref{relpart-FS-conservation}, which is what one obtains with the Fock--Stueckelberg theory. Thus, the Fock--Stueckelberg approach leads to a relaxation of first-class constraints by adding new degrees of freedom, which generally can be seen as the $w$ multipliers [e.g., as in Eq. \eref{mechFS1}], but, in certain cases, these new variables can be identified with constants of motion (promoted to physical conserved fields), as it was discussed in Sec. \ref{sec:constants}, and as we see here in the case of the relativistic particle, where $\mathcal{E}$ may be related to the particle's rest energy.

\subsubsection{The work of Fock and Stueckelberg}
The above formalism is a direct generalization of the original work of Fock and Stueckelberg. In \cite{Fock}, Fock considered a charged relativistic particle and the use of proper time as an independent variable in the particle's action principle. In that work, it was noted that, in the usual formulation of the relativistic particle's action principle [which can be generalized to the action in Eqs. \eref{mechaction3} and \eref{mechaction4}], proper time is not an independent variable (as it depends on the path of the particle) but, with the introduction of a new action [which can be generalized to the action in Eq. \eref{mechaction5}], one could use an independent quantity that coincided with proper time on a solution to the equations of motion. The connection with the usual treatment of the relativistic particle proceeded by fixing a constant of motion that would recover the usual time-reparametrization invariant formalism. This is precisely what is done in Eq. \eref{back-to-Jacobi}, which is a generalization of the equation $\dot{x}^2+\dot{y}^2+\dot{z}^2-c^2\dot{t}^2 = \text{const.}\to-c^2$ that is discussed in Fock's article.\footnote{To obtain $\dot{x}^2+\dot{y}^2+\dot{z}^2-c^2\dot{t}^2 = \text{const.}\to-c^2$ from Eq. \eref{back-to-Jacobi}, we set $g_{ab}$ to be the Minkowski metric with mostly plus signature, $V\equiv0$, and $\mathcal{E}=-mc^2/2$.} Fock also discussed the implications of the use of an independent proper time for the Dirac equation.

Similarly, Stueckelberg proposed in \cite{Stueck,Stueck1,Stueck2} a ``new relativistic mechanics'' by positing a new action principle for the relativistic particle. The proposed action was a particular case of Eq. \eref{mechaction5}, and Stueckelberg stressed that the new formulation could be straightforwardly quantized \`{a} la Schr\"odinger, and it could also be used to describe the creation and annihilation of particles not only in the quantum theory but also in the classical theory [due to the fact that the trajectories that can be derived from Eq. \eref{mechaction5} need not be always timelike or null].\footnote{The works \cite{Stueck,Stueck1,Stueck2} of Stueckelberg have a conceptual similarity with but are distinct from the other famous Stueckelberg method \cite{StueckTrick,StueckTrick1,StueckTrick2}: that of introducing a scalar field to preserve gauge invariance in a massive Abelian gauge theory of the Yang--Mills type.} The value of the Hamiltonian (or, in the particular case of a free particle, the particle’s mass) was to be understood as an integration constant associated to each worldline (an idea which would later make a comeback concerning the cosmological constant in unimodular gravity, and possibly any constant of Nature in quantum gravity \cite{Magueijo1,Magueijo2,Magueijo3,Magueijo4,Magueijo5,Magueijo6,Magueijo7}, cf. Secs. \ref{sec:constants} and \ref{sec:reappear}).

The work of Fock and Stueckelberg was developed in the context of the early days of relativistic quantum theory (and ultimately quantum field theory). It has inspired many subsequent applications, including the relativistic quantum mechanics which was further developed by Horwitz, Piron and others \cite{HPStueck1,HPStueck2,HPStueck3}. In the context of quantum gravity and gauge theories, their work is seldom acknowledged. Although Fock and Stueckelberg only considered a particular gauge theory (the relativistic particle with its local worldline reparametrizations) and they did not employ the modern tools of canonical gauge theory and first-class constraints discussed in the present article,\footnote{This is historically understandable. When their works were published, a general theory of first-class constrained Hamiltonian systems did not seem to be generally known, despite early works of Rosenfeld \cite{Rosenfeld,SalisburyRosenfeld,RBD}, and it would only rise to prominence with the articles of Bergmann and Dirac \cite{Dirac1,Dirac2,Dirac3,DiracGravity,Bergmann}.} the modern reader can clearly grasp that the essence of their approach is gauge-theoretical, in the sense of producing a new theory from a parent gauge theory, as we have explained in the previous sections.

\subsection{\label{sec:DiracEM}Electrodynamics and ``charge without charge''}
Another interesting application of the Fock--Stueckelberg mechanism is in the archetypical gauge theory: electromagnetism. Let us consider a four-dimensional Minkowski spacetime with metric signature $(-,+,+,+)$, and let us now use Latin indices to indicate the components of spacetime tensors, not the multiplier fields as in previous sections. We use units in which the speed of light is $c=1$ for simplicity. We also no longer adopt DeWitt's compact notation for clarity (i.e., repeated indices are tacitly summed over their discrete values but no accompanying spatial integration is implicit). The Maxwell Lagrangian reads
\eq{\label{eq:Maxwell}
\mathcal{L} &= -\frac14F^{ab}F_{ab} \ ,\\
F_{ab} &= \partial_{a}A_{b}-\partial_{b}A_{a} \ , 
}
where the $z$ fields are to be understood as the canonical pairs of spatial components of $A_{i}$ ($i = 1,2,3$) and their conjugate momenta [obtained after a Legendre transform of the variables in Eq. \eref{Maxwell}], whereas $A_0$ plays the role of a multiplier that enforces the Gauss law (see, e.g., \cite{Hanson,Kiefer:book,SalisburyMomentum} for the Hamiltonian formulation of Maxwell's theory).

Let us consider the gauge condition
\eq{\label{eq:Dirac-gauge}
\chi = \frac{1}{2}(A^{a}A_{a}+k^2)\approx0 \ , 
}
with $k$ being a constant. This is a noncanonical gauge, as it is meant to fix the multiplier $A_0$. We obtain $A_0 = \pm\sqrt{A_1^2+A_2^2+A_3^2+k^2}$, which is well defined (in the sense of yielding a real $A_0$ and not introducing any new restrictions on the other components of the four-vector potential) if $k^2\geq0$ (corresponding to a timelike or null four-vector potential) \cite{NambuEM}. This gauge condition is interesting because, given the gauge transformation $A_{a} = \tilde{A}_{a}+\partial_{a}\varepsilon$, where $\tilde{A}_{a}$ is the four-vector potential in some other arbitrary gauge, we find
\eq{
g^{ab}\left(\del{\varepsilon}{x^{a}}+\tilde{A}_{a}\right)\left(\del{\varepsilon}{x^{b}}+\tilde{A}_{b}\right) \approx -k^2 \ ,
}
which is equivalent to the Hamilton--Jacobi equation for a charged relativistic particle [cf. Eq. \eref{HJ-Dirac-gauge}] if we identify $\varepsilon\equiv -S/e$, $k = mc/e$ (with $c=1$ in the adopted units), and $\tilde{A}_a\leftrightarrow A_a$ \cite{Dirac,NambuEM}. Thus, this gauge choice is accessible only if the Hamilton--Jacobi equation for the charged particle has a solution.

This equivalence with the dynamics of charged particles motivated Dirac \cite{Dirac} to consider that the Maxwell theory could be the ``wrong'' classical theory to be quantized, given the lack of details about the structure of the electron and the ensuing complications that researchers faced in order to describe the electron's self-energy in point-charge models\footnote{This was eventually addressed by renormalization, which Dirac considered unappealing \cite{Dirac}.} or in models in which the charge was not localized, which proved to be intractable at the time. Rather, Dirac posited, one ought to seek a theory that is simple in the sense that it makes no assumptions about the structure of electrons, so as to lead to a simpler quantization. This is similar to Stueckelberg's motivation to propose a ``new relativistic mechanics'' with a more direct quantization procedure.

As the gauge invariance of electromagnetism leads to extra variables that do not have physical significance,\footnote{They are inaccessible ``generalized reference frames.''} one should break this invariance, according to Dirac, and use the extra variables to describe the sought-after charged particles. This led to Dirac's proposal of a new classical Lagrangian for electrodynamics:\footnote{There were of course previous works by other authors that modified the Maxwell Lagrangian in order to break gauge invariance. See, for instance, \cite{Fermi,Proca}.}
\eq{\label{eq:DiracEM-Lagrangian}
\mathcal{L} &= -\frac14F^{ab}F_{ab}-\frac{w}{2}(A^{a}A_{a}+k^2) \ ,
}
which is readily understood to be a particular case of the general Fock--Stueckelberg approach [cf. Eq. \eref{FS-action}]. The Fock--Stueckelberg Lagrangian in Eq. \eref{DiracEM-Lagrangian} leads to the field equations
\eq{
\partial_{a}F^{ab} = wA^{b} =: -J^{b} \ ,
}
with $\partial_{a}J^{a}=0$ due to the antisymmetry of $F_{ab}$. The appearance of the four-current $J^{a}$ signals the presence of new degrees of freedom with respect to the parent theory. In particular, we obtain a modification of the vacuum Gauss law:
\eq{
{\rm div}\vec{E} = -wA^0 =: J^0 =:\rho \ .
}
As discussed in the previous sections, this constraint relaxation is a characteristic feature of the Fock--Stueckelberg theory. For the case of electromagnetism, we obtain ``electrons without electrons'' or ``charge without charge.'' Furthermore, Dirac emphasized that his new electrodynamics is equivalent to Maxwell's theory if the $w$ multiplier vanishes (absence of charges). As discussed above, this is a general feature of the Fock--Stueckelberg approach: if $w$ is zero on shell, then we recover the parent theory, which, in this case, is vacuum electromagnetism.

Dirac's electrodynamics inspired many further works (e.g., \cite{post,post1,post11,post2,post3}),\footnote{It is worthwhile to mention that Nambu \cite{NambuEM} verified that the corresponding quantum field theory and $S$-matrix of Dirac's new electrodynamics with a timelike four-vector potential (with $k^2>0$) agree with the usual results of quantum electrodynamics at least up to the lowest orders in perturbation theory. More precisely, Nambu considered the Lagrangian $-\frac14F^{ab}F_{ab}-j_{a}A^{a}$, where $j^a$ is a conserved current, and the gauge condition is tacitly already satisfied; i.e., $A_0$ is defined as $A_0 \Def \pm\sqrt{A_1^2+A_2^2+A_3^2+k^2}$. This corresponds to Dirac's new electrodynamics in the presence of some four-current $j^a$ and with the gauge condition already imposed at the level of the action [similarly to Eq. \eref{FS-action-1}].} including an extension to non-Abelian gauge theories in \cite{post11}, and the idea was recently reincarnated in \cite{Kaplan2}, which is part of the series of works \cite{Kaplan,Kaplan1,Kaplan2}, where the argument that the modified theory leads to a more straightforward quantization is again recorded. However, both Stueckelberg and Dirac make it clear that one is producing a different \emph{classical} theory which is to be quantized, rather than arguing that the modified theory follows from a consistent quantization or concerns the ``classical equations of motion of quantized gauge theories'' (as in \cite{Kaplan1,Kaplan2}). The Fock--Stueckelberg approach does not concern the consistency of quantization methods of gauge theories,\footnote{One such method is the powerful BRST cohomology \cite{HT:book}, which does not require that a gauge be fixed prior to quantization in order to achieve a nonambiguous Hamiltonian.} but it rather offers an ingenious way to classically add new degrees of freedom by breaking the original gauge symmetry.\footnote{Of course, Nature is presumably ``quantum first,'' and classical behavior is relegated to an approximate limit. Even so, this does not imply that, in order to be well defined, the fundamental quantum theory must break gauge symmetries, which would then only emerge in particular cases. We will further discuss this in Sec. \ref{sec:conclusions}.}

Dirac's treatment can be generalized by considering the Lagrangian
\eq{
\mathcal{L} &= -\frac14F^{ab}F_{ab}+A_{a}j^{a}-w \chi(A) \ ,
}
where $\chi(A)$ is a generic gauge condition on the four-vector potential, and $j^{a}$ is a ``real'' four-current (associated with local degrees of freedom already present in the parent theory). The corresponding Fock--Stueckelberg field equations are
\eq{
\partial_{a}F^{ab} = -j^b+w\del{\chi}{A_b} =: -(j^b+J^b) \ ,
}
which, in particular, imply a modification of the parent-theory Gauss-law constraint:
\eq{
{\rm div}\vec{E} = j^0-w\del{\chi}{A_0} =: j^0+J^0 =:\rho \ .
}
An example that was considered in \cite{Kaplan2} is the Weyl gauge $\chi(A) = A_0$. In this case, we have $J^b = (-w,\vec{0})$, so that there is a violation of the parent-theory Gauss law but not of the Amp\`{e}re law. The physical interpretation of the Fock--Stueckelberg effective current in this case is less clear than in Dirac's original treatment connected with the charged particle's Hamilton--Jacobi equation.

\subsection{\label{sec:gug-0}Generalized unimodular gravity and ``matter without matter''}
The Fock--Stueckelberg approach can also be applied to general relativity. In what follows, as in the relativistic particle and electromagnetic cases, repeated indices are to be summed over their discrete values but there is no implicit spatial integration. As before, we consider a spacetime with a mostly positive metric signature but now spacetime indices are denoted by Greek letters, whereas spatial indices are denoted by Latin letters.\footnote{This notation is not to be confused with the $\mu$ and $a$ indices used in Secs. \ref{sec:gauge-theory} and \ref{sec:FS-gen}.} We also adopt units in which the speed of light $c$ and Newton's constant $G$ are both equal to one, $c=G=1$.

\subsubsection{\label{sec:canonicalGR}Canonical general relativity}
Given a spacelike foliation of spacetime, the spacetime metric with components $g_{\mu\nu}$ and determinant $g\Def\det{g_{\mu\nu}}$ can be decomposed as \cite{ADM,Kiefer:book}\footnote{Obviously, there is an abuse of notation in Eq. \eref{3p1decomp}, as we use the components of matrices and vectors to denote the matrices and vectors themselves.}
\eq{\label{eq:3p1decomp}
g_{\mu\nu} &= \left(\begin{matrix}N_aN^a-N^2 & N_b\\ N_b & h_{ab}\end{matrix}\right) \ ,\\ \ g^{\mu\nu} &= \left(\begin{matrix}-\frac{1}{N^2} & \frac{N^a}{N^2}\\ \frac{N^a}{N^2} & h^{ab}-\frac{N^aN^b}{N^2}\end{matrix}\right)
}
where the spatial indices are lowered and raised by the spatial metric $h_{ab}$, which has determinant $h\Def\det h_{ab} = |g|/N^2$ and inverse $h^{ab}$. The quantities $N, N^a$ are respectively the lapse function (taken to be always positive) and the shift vector. These are the variables used in the Arnowitt--Deser--Misner (ADM) formulation of general relativity \cite{ADM}.

In this case, the $z$ fields of the formalism presented in Secs. \ref{sec:gauge-theory} and \ref{sec:FS-gen} correspond to matter fields and the components $h_{ij}$ of the spatial metric, whereas $N,N^a$ act as multipliers.\footnote{This implies that the constraints given in Eqs. \eref{GR-constraints} are treated as primary constraints in accordance with the formalism of Secs. \ref{sec:gauge-theory} and \ref{sec:FS-gen}. However, we can evidently include the momenta conjugate to $N,N^a$ in the theory, so as to treat the lapse and shift as ordinary non-multiplier fields. In this case, their momenta must vanish as primary constraints, whereas the constraints in Eqs. \eref{GR-constraints} become secondary \cite{Kiefer:book,Pons}.} The first-class constraints associated with these multipliers are, respectively,\footnote{We disregard the presence of a cosmological constant in Eqs. \eref{GR-constraints} because, as we shall see, it can be recovered in the corresponding Fock--Stueckelberg theory.}
\eq{\label{eq:GR-constraints}
\phi_{\perp} &= 16\pi G_{abcd}\ p^{ab}p^{cd} -\frac{\sqrt{h}}{16\pi }{^{(3)}}{R}+\sqrt{h}\rho = 0 \ , \\
\phi_a &= -2\mathrm{D}_b\tensor{p}{_a^b}+\sqrt{h}J_a = 0 \ , 
}
where $p^{ab}$ is the canonical momentum conjugate to the spatial metric (and it is a spatial tensor density of weight one); $^{(3)}R$ and $\mathrm{D}_a$ are, respectively, the Ricci scalar and the covariant derivative on a leaf of the foliation; and $\rho$ and $J_a$ respectively correspond to the matter energy density and ``Poynting vector'' \cite{Kiefer:book,ADM,Hanson,MTW,DeWittTrilogy1}. The inverse metric on the gravitational configuration space reads
\eq{\label{eq:DeWitt-inv-metric}
G_{abcd} = \frac{1}{2\sqrt{h}}\left(h_{ac}h_{bd}+h_{ad}h_{bc}-h_{ab}h_{cd}\right) \ ,
}
whereas the gravitational configuration-space metric $G^{abcd}$ satisfies $G^{abcd}G_{cdef} = (\delta^a_e\delta^b_f+\delta^a_f\delta^b_e)/2$. Moreover, the Hamiltonian field equations can be used to write the momentum $p^{ab}$ as
\eq{\label{eq:GR-momentum}
p^{ab} = \frac{1}{32\pi N}G^{abcd}\left(\dot{h}_{cd}-\mathrm{D}_c N_d-\mathrm{D}_d N_c\right) \ .
}
The first-class algebra of general relativity is given by the Poisson brackets (suppressing the temporal and field dependence for brevity)
\begin{align}
\{\phi_{\perp}(\x),\phi_{\perp}(\y)\} \!&=\! \delta_{,a}(\x,\y)\left[h^{ab}(\x)\phi_b(\x)+h^{ab}(\y)\phi_b(\y)\right],\notag\\
\{\phi_{a}(\x),\phi_{\perp}(\y)\} &= \phi_{\perp}(\x)\delta_{,a}(\x,\y)\,,\label{eq:GR-algebra}\\
\{\phi_a(\x),\phi_b(\y)\} &=\phi_b(\x)\delta_{,a}(\x,\y)+\phi_a(\y)\delta_{,b}(\x,\y)\,,\notag
\end{align}
where $\delta_{,a}(\x,\y)$ denotes a partial derivative of the Dirac delta distribution with respect to $x^a$.

\subsubsection{\label{sec:gug}Fock--Stueckelberg approach to general relativity}
An example of a partial non-canonical gauge fixing of the kind given in Eq. \eref{FS-gf-main} is given by the condition \cite{we}
\eq{\label{eq:gug-gauge}
\chi\Def N - \Lambda(h) \approx 0 \ ,
}
which fixes the lapse to be a function of the determinant of the spatial metric, $N\approx\Lambda(h)$. We assume that $\Lambda(h)$ is always positive. In this case, the constraints of the Fock--Stueckelberg theory are $\chi, \phi_{\perp}+w,$ and $\phi_a$ [cf. Eqs. \eref{FS-constraints}], and thus the first-class Hamiltonian constraint $\phi_{\perp}$ of general relativity, which results from the variation of the action with respect to $N$, is relaxed. Its value can now be arbitrary, as it simply fixes the value of the new multiplier field $w$, which is to be interpreted as an effective kind of matter.\footnote{This is analogous to what is accomplished by Synge's G-method \cite{Synge}. See the comments in \cite{Ellis}.}

From the variation of the Fock--Stueckelberg action \eref{FS-action}, we obtain the Hilbert energy-momentum tensor of the effective matter,\footnote{In the functional variation $\delta \mathcal{F} = \varepsilon\left.\Del{}{\varepsilon}\right|_{\varepsilon=0}\mathcal{F}(\tau,\x;g_{\mu\nu}+\varepsilon \eta_{\mu\nu}] =: \varepsilon\int\D^3x\ \frac{\delta \mathcal{F}}{\delta g_{\mu\nu}}\eta_{\mu\nu}(\tau,\x)$ (cf. Footnote \ref{foot:fder}), as the metric and its (arbitrary) variation $\eta_{\mu\nu}$ are symmetric tensors, we must symmetrize the functional derivative $\delta \mathcal{F}/\delta g_{\mu\nu}$ with respect to the $\mu,\nu$ indices. For example, from Eq. \eref{inverse-metric-var}, we find $\delta g^{\mu\nu}(\tau,\mathbf{x})/\delta g_{\rho\sigma}(\tau,\mathbf{y}) = -\delta(\mathbf{x},\mathbf{y})(g^{\mu\rho}g^{\sigma\nu}+g^{\mu\sigma}g^{\rho\nu})_{(\tau,\mathbf{x})}/2$.}
\eq{
T^{\mu\nu}_{\rm eff} \Def \frac{2}{\sqrt{|g|}}\frac{\delta}{\delta g_{\mu\nu}}\int\D^4 x\,\left(-w\chi\right) \ ,
}
which can be computed by using the relation
\eq{\label{eq:inverse-metric-var}
\delta g^{\mu\nu}=-g^{\mu\rho}g^{\sigma\nu}\delta g_{\rho\sigma}
}
to find
\eq{
\delta N &= \delta\left[(-g^{00})^{-\frac12}\right] =-\frac{N^3}{2}g^{0\mu}g^{\nu0}\delta g_{\mu\nu} \ .
}
The relation $\delta h = h h^{ij}\delta h_{ij} = h h^{ij}\delta_i^{\mu}\delta_j^{\nu}\delta g_{\mu\nu}$ is also useful. With this, the on-shell result is seen to be the energy-momentum tensor of a perfect fluid with four-velocity $n^{\mu} = (1/N,-N^i/N)=-g^{0\mu}N$ with $N\approx\Lambda$, which is the normal vector to the spatial hypersurfaces, and with energy density
\eq{\label{eq:gug1}
\mathscr{E} \approx \frac{w}{\sqrt{h}} \ ,
}
and pressure
\eq{\label{eq:gug2}
\mathscr{P} \approx 2\frac{\D\log \Lambda}{\D\log h}\mathscr{E} =: \mathscr{W}\mathscr{E} \ .
}
What can we conclude about the function $\Lambda(h)$ that defines the gauge fixing? And what about the new multiplier field $w$ that characterizes the effective fluid? First, notice that the constraints $\chi\approx0,\phi_{\perp}+w\approx0,\phi_a\approx0$ are the `primary' constraints in the Fock--Stueckelberg theory, as they follow from the Fock--Stueckelberg action itself. The constraints $\chi\approx0$ and $w\approx-\phi_{\perp}$ fix the multipliers $N$ and $w$, respectively, at all times, and their preservation in time fixes the corresponding derivatives of the multipliers. On the other hand, we must proceed with the analysis of the constraint conservation (\`{a} la Rosenfeld--Bergmann--Dirac \cite{Rosenfeld,SalisburyRosenfeld,Dirac3,RBD,HT:book,Bergmann}), specifically with respect to the (``unaltered'') spatial diffeomorphism constraints $\phi_a$ to determine whether new (`secondary') constraints appear in the theory and whether the Fock--Stueckelberg theory defined from the parent-theory gauge choice given in Eq. \eref{gug-gauge} is consistent. We can use the field equations \eref{FS-eom} together with the first-class algebra given in Eqs. \eref{GR-algebra} to find {(suppressing the dependence of the constraints on the fields)}
\begin{align}
0&\approx\dot{\phi}_a(\tau,\x)\approx \int\!\D^3y\, \{\phi_a(\tau,\x),\Lambda(h(\tau,\y))\phi_{\perp}(\tau,\y)\}\notag\\
&\approx -2\del{}{x^a}\left(w\del{\Lambda}{\log h}\right)\notag\\
&=-\del{}{x^a}\left(w\Lambda\mathscr{W}\right)\ . \label{eq:secondary}
\end{align}
This is a new (secondary) constraint that is not redundant with any of the primaries. We must then impose it at all times.
For this, it is useful to compute [cf. Eq. \eref{GR-momentum}]
\begin{align}
-\dot{w}(\tau,\x)&\approx\dot{\phi}_{\perp}(\tau,\x)\notag\\
&\approx\int\D^3y\, \{\phi_{\perp}(\tau,\x),\Lambda(h(\tau,\y))\phi_{\perp}(\tau,\y)\}\notag\\
&\ \ \ + \int\D^3y\, \{\phi_{\perp}(\tau,\x),N^a(\tau,\y)\phi_{a}(\tau,\y)\}\notag\\
&\approx \left[\frac{w\dot{\Lambda}}{\Lambda}-w\mathscr{W}\del{N^a}{x^a}\right.\\
&\ \ \ \left.-\frac{w}{\Lambda}N^a\del{\Lambda}{x^a}-\del{}{x^a}\left(N^aw\right)\right]_{(\tau,\x)} \,,\notag
\end{align}
which is equivalent to
\eq{\label{eq:sec-Ham-const}
\Del{}{\tau}\left(w\Lambda\right)-\del{}{x^a}\left[N^a w\Lambda\left(1+\mathscr{W}\right)\right]\approx0 \ .
}
With Eq. \eref{sec-Ham-const}, we can compute
\eq{\label{eq:tertiary}
0&\approx -\Del{}{\tau}\del{}{x^a}\left(w\Lambda\mathscr{W}\right)\\
&=-\del{}{x^a}\left\{\dot{\mathscr{W}}w\Lambda+\mathscr{W}\del{}{x^b}\left[N^bw\Lambda(1+\mathscr{W})\right]\right\} \ ,
}
which is a new (`tertiary') constraint that can be used to fix the shift vector $N^a$ if it is not redundant with the other constraints.\footnote{It is important to notice that constraints are relations on phase space that can be used to restrict either the initial values of the fields and their momenta or the multipliers, and they are enforced for every solution of the theory. Thus, a constraint of the form given in Eq. \eref{tertiary} can in principle be used to fix $N^a$ for every admissible choice of initial values of the fields and their momenta.} In general, the solution to the secondary constraint in Eq. \eref{secondary} is
\eq{\label{eq:secondary2}
w\Lambda\mathscr{W} \approx \mathscr{F}(\tau) \ ,
}
where $\mathscr{F}(\tau)$ is a function solely of time. In the cases in which $\mathscr{F}(\tau)\neq0$ at all times, then the tertiary constraint becomes
\eq{\label{eq:tertiary-2}
\del{}{x^a}\left\{(1+\mathscr{W})\del{N^b}{x^b}+\left(\Del{}{\tau}-N^b\del{}{x^b}\right)\log\mathscr{W}\!\right\} \!\approx0\,,
}
which is consistent with
\eq{\label{eq:tertiary3}
(1+\mathscr{W})\del{N^b}{x^b}+\left(\Del{}{\tau}-N^b\del{}{x^b}\right)\log\mathscr{W}\!\approx\!\Del{\log\mathscr{F}}{\tau}\,.
}
This generally fixes the shift vector $N^a$ (it cannot be left arbitrary), thereby also breaking the three-dimensional diffeomorphism symmetry on the spatial hypersurfaces. The exceptions are the cases in which Eq. \eref{tertiary} is independent of $N^a$. These are: (1) $w$ is everywhere zero on shell (in which case the Fock--Stueckelberg theory is equivalent to general relativity); (2) if $\mathscr{W}\equiv0$, which corresponds to an effective dust matter [cf. Eq. \eref{gug2}]; (3) if $\mathscr{W}\equiv-1$, which corresponds to an effective cosmological constant [cf. Eq. \eref{gug2}] (this is the case of unimodular gravity). In all three cases, the tertiary constraint is trivialized (i.e., it does not lead to any new restrictions on the fields and their momenta nor the multipliers). In cases (1) and (2), we also have $\mathscr{F}(\tau)\approx0$ at all times. For case (3), we find that Eqs. \eref{sec-Ham-const} and \eref{secondary2} determine that $\mathscr{F}$ is a spacetime constant. In this case, we obtain [cf. Eqs. \eref{gug1}, \eref{gug2} and \eref{secondary2}]
\eq{\label{eq:gug-uni}
\Lambda &= \frac{\Lambda_0}{\sqrt{h}} \ , \ 
w \approx \frac{\mathscr{F}}{\Lambda_0}\sqrt{h} \ , \\
\mathscr{E} &\approx \frac{\mathscr{F}}{\Lambda_0} = \text{const.} =:\mathscr{E}_0 \ ,
}
where $\Lambda_0$ and $\mathscr{F}$ are spacetime constants, so as to yield the energy density $\mathscr{E}$ that corresponds to the constant energy density of a cosmological constant. Furthermore, the gauge condition $N\approx\Lambda = \Lambda_0/\sqrt{h}$ is equivalent to the determinant condition $|\det g_{\mu\nu}| = \Lambda_0^2$. By setting $\Lambda_0 = 1$, we recover the usual unimodular gravity condition. Thus, unimodular gravity can be seen as a particular case of the Fock--Stueckelberg approach to general relativity.

By keeping a more general $\Lambda(h)$, we obtain a theory that can be called ``generalized unimodular gravity,'' which was discussed in \cite{we} and further developed with different applications in \cite{Barv, Barv1, Kam-Tron-Ven}.\footnote{In particular, the article \cite{Barv} discussed the constraint algebra in some detail.} Depending on the choice of $\Lambda(h)$ [i.e., the choice of partial gauge fixing in Eq. \eref{gug-gauge}] and the consistency of the constraint preservation equations \eref{secondary} and \eref{tertiary}, different residual gauge symmetries might remain (cf. Sec. \ref{sec:gf}). Whereas the residual gauge symmetries of unimodular gravity correspond to the group of volume/area-preserving diffeomorphisms [and the shift vector $N^a$ is not fixed by Eq. \eref{tertiary}], the residual gauge symmetries of the generalized unimodular theory may be more involved (see \cite{we,Barv}). One might also complement the condition in Eq. \eref{gug-gauge} with an extra gauge condition on the shift vector [e.g., $N^a\approx\Lambda^a(h)$] and subsequently analyze the consistency of the Fock--Stueckelberg constraints.

In this way, the application of the Fock--Stueckelberg approach to general relativity naturally leads to new degrees of freedom [coming from the $w$ multiplier(s)] that can be effectively identified with different kinds of matter, whose form depends on the chosen gauge condition(s). This effective matter appears in addition to the matter content that was already present in the parent theory. If the parent theory is purely gravitational, the corresponding Fock--Stueckelberg theory will feature a gravitational field and an effective fluid. Similarly to Dirac's electrodynamics, we thus have ``matter without matter'' in the Fock--Stueckelberg theory.

In \cite{Kaplan1}, this emergence of effective matter from a relaxation of the first-class constraints in general relativity was rediscovered. It is also important to mention different approaches across the literature, which yield similar results or employ a similar reasoning to the Fock--Stueckelberg approach. The treatment of constrained Hamiltonian systems via the ``unfree gauge symmetry'' formalism was discussed in the series of papers \cite{unfree,unfree1,unfree2,unfree3}. The essential idea is to consider gauge transformations where $\varepsilon$ (cf. Sec. \ref{sec:gauge-transf}) are restricted by additional conditions, namely, they must satisfy a set of partial differential equations. This restriction of the gauge transformations leads to a new theory, just as in the case of the Fock--Stueckelberg method. Indeed, the simplest example is again unimodular gravity, where the corresponding equations are simply the conditions for transverseness, $ \nabla_{\mu}\varepsilon^{\mu} = 0$ (with $\mu$ a spacetime index). Moreover, unimodular gravity was analyzed from the point of view of restricted gauge theories in \cite{Barv-Nest}. Finally, in the so-called TDiff theories (see, e.g., \cite{Tdiff}), the matter part of the action is invariant only with respect to the transverse diffeomorphisms, which leads to a weaker version of unimodular gravity. Some models of similar theories were considered in \cite{Tdiff1}, where the details of the breaking of diffeomorphism invariance were discussed. The reader is also referred to \cite{Percacci} for a rather broad analysis of gravity theories with different sets of constraints or different structures of gauge algebras. All in all, the Fock--Stueckelberg approach can be seen as one of the oldest and simplest of these approaches where the gauge symmetry is modified to yield a new theory from a parent.

\subsubsection{Homogeneous cosmology}
One of the most useful applications of generalized unimodular gravity is in cosmology. This corresponds to an application of the Fock--Stueckelberg approach to a toy model of general relativity, such as the Friedmann--Lama\^{i}tre--Robertson--Walker (FLRW) model with a scalar field $\varphi$. For a spatially flat model, the spacetime metric is given by the line element
\eq{
\D s^2 = -N^2(\tau)\D\tau^2+a^2(\tau)\D\mathbf{x}^2 \ ,
}
where $N(\tau)$ is the lapse function, the shift vector $N^a$ and spatial diffeomorphism constraint are absent due to the homogeneity assumption, $a(\tau)$ is the scale factor of the universe, and $\D\mathbf{x}^2$ is the line element of three-dimensional Euclidean space. The Fock--Stueckelberg Lagrangian reads\footnote{In this homogeneous setting, the spatial integration in the action leads to an overall factor of $\mathfrak{L}^3$, which is a fiducial spatial volume, multiplying the Lagrangian, $\D\tau\mathfrak{L}^3 L$. The arbitrary length scale $\mathfrak{L}$ can be removed by the redefinitions $a\to a/\mathfrak{L}, N\to N/\mathfrak{L}, t\to \mathfrak{L}t$ \cite{Redef}, which lead to the Lagrangian in Eq. \eref{FS-La-cosm}. The gauge condition $\chi$ is defined with respect to these rescaled variables. We have also changed units to set $G=1/(8\pi)$.}
\eq{\label{eq:FS-La-cosm}
L &= -3\frac{a\dot{a}^2}{N}+\frac{a^3\dot{\varphi}^2}{2N}-Na^3\mathcal{V}(\varphi)-w\chi\\
&=p_a\dot{a}+p_{\varphi}\dot{\varphi}-N\phi_{\perp}-w\chi \ .
}
In the absence of the spatial diffeomorphism constraints $\phi_a$, the only Fock--Stueckelberg constraints are $\chi = N-\Lambda\approx0$ and $\phi_{\perp}+w\approx0$, both of which fix the multipliers $N$ and $w$ at all times. The Hamiltonian constraint \cite{DeWittTrilogy1,Kiefer:book,MTW},
\eq{\label{eq:cosm-Ham-constraint}
\phi_{\perp} = -\frac{p_a^2}{12a}+\frac{p_{\varphi}^2}{2a^3}+a^3\mathcal{V}(\phi) \ ,
}
is then relaxed, and the analysis that led to the secondary and tertiary constraints in Eqs. \eref{secondary} and \eref{tertiary} need be performed no longer. The model is thus mechanical and, in fact, it corresponds to a relativistic particle in a curved background, so that we can use the formalism of Secs. \ref{sec:mech} and \ref{sec:relpart}. The analogue of Eq. \eref{sec-Ham-const} leads us to
\eq{\label{eq:cosm-w}
\Del{}{\tau}\left(w\Lambda\right)\approx0 \ ,
}
which implies $w\approx w_0/\Lambda$, where $w_0$ is a spacetime constant.

For the purposes of cosmology, it suffices to choose a gauge condition for which the effective fluid equation-of-state parameter $\mathscr{W}$ is a spacetime constant, which implies [cf. Eqs. \eref{gug1}, \eref{gug2} and \eref{cosm-w}]
\eq{\label{eq:gug-fluid-cosm}
\Lambda &= \Lambda_0 a^{3\mathscr{W}}  \ , \ 
w \approx \frac{w_0}{\Lambda_0}a^{-3\mathscr{W}} \ , \\
\mathscr{E} &\approx \frac{w_0}{\Lambda_0}a^{-3(\mathscr{W}+1)} =: \mathscr{E}_0a^{-3(\mathscr{W}+1)}  \ ,
}
with $\Lambda_0$ being a spacetime constant. The quantities in Eq. \eref{gug-fluid-cosm} reduced to those in Eq. \eref{gug-uni} for $\mathscr{W}=-1$, as they should.

From the Fock--Stueckelberg Lagrangian in Eq. \eref{FS-La-cosm}, we obtain the equations
\eq{\label{eq:eom-cosm}
0&=3\left(\frac{\dot{a}}{Na}\right)^2-\frac12\left(\frac{\dot{\varphi}}{N}\right)^2-\mathcal{V}-\frac{w_0}{\Lambda_0}a^{-3(\mathscr{W}+1)} \ , \\
0&=\frac{2}{N^2}\frac{\ddot{a}}{a}+\frac{1}{N^2}\left(\frac{\dot{a}}{a}\right)^2-2\frac{\dot{N}\dot{a}}{N^3 a}\\
&\ \ \ +\frac{\dot{\varphi}^2}{2N^2}-\mathcal{V}-\frac{w_0\mathscr{W}}{\Lambda_0a^{3(\mathscr{W}+1)}}\ , \\
0&=\ddot{\varphi}+3\frac{\dot{a}\dot{\varphi}}{a}-\frac{\dot{N}\dot{\varphi}}{N}+N^2\del{\mathcal{V}}{\varphi} \ ,
}
where $N\approx\Lambda$. Notice that the terms proportional to $w_0$ constitute a modification of the standard Friedmann equations, which can indeed be interpreted as contributions of an effective fluid with a constant equation-of-state parameter $\mathscr{W}$. Thus, in particular, by assuming $w_0/\Lambda_0>0$, we have: effective dust (dark matter) for $\mathscr{W}=0$, for which we recover the gauge $N \approx \Lambda_0$;\footnote{A generalized unimodular gravity with effective dust matter and $N = 1$ gauge was also considered in the book \cite{Burlan} and the series of articles cited therein.} an effective cosmological constant (dark energy) for $\mathscr{W}=-1$ (which corresponds to the unimodular gauge condition in the parent theory); an effective curvature contribution for $\mathscr{W}=-1/3$; and effective radiation contribution for $\mathscr{W}=1/3$.

\subsubsection{\label{sec:GRconstants}Time from constants of Nature}
As in Secs. \ref{sec:constants} and \ref{sec:time-conjugate-constant}, there are certain cases in which time can be seen as the canonical conjugate to a constant of Nature also in the Fock--Stueckelberg approach to general relativity (generalized unimodular gravity). For example, in the case of homogeneous cosmology with a constant effective equation-of-state parameter, the Fock--Stueckelberg equations of motion are [cf. Eq. \eref{FS-eom}]
\eq{\label{eq:cosm-eom-constant}
\dot{f} = \{f,\Lambda_0 a^{3\mathscr{W}}\phi_{\perp}\}
}
for any phase-space function $f\equiv f(a,p_a;\varphi,p_{\varphi})$, and the Fock--Stueckelberg Hamiltonian is conserved but not necessarily zero (the constraint is relaxed),
\eq{\label{eq:cosm-eom-constant-ham}
\!\Lambda_0 a^{3\mathscr{W}}\phi_{\perp} \approx-w_0 \,,
}
due to the Fock--Stueckelberg constraint $\phi_{\perp}+w\approx0$ [cf. Eqs. \eref{cosm-Ham-constraint}, \eref{gug-fluid-cosm} and \eref{eom-cosm}] and the fact that $w_0$ is a spacetime constant that need not vanish (if it vanishes, then the theory is identical to the ordinary homogeneous cosmology derived from general relativity). 

Are the results in Eqs. \eref{cosm-eom-constant} and \eref{cosm-eom-constant-ham} also obtained by promoting a constant of Nature to a physical field conjugate to time? Instead of obtaining an effective fluid via the Fock--Stueckelberg approach to general relativity with a scalar field, let us consider ordinary general relativity coupled to a scalar field and a fluid with energy density given by $\mathscr{E}=\mathscr{E}_0a^{-3(\mathscr{W}+1)}$ (i.e., the fluid is now already present in the parent theory but it is treated as a ``potential term'' dependent on $a$). The equations of motion of the corresponding homogeneous cosmology are
\eq{
\dot{f} &= \{f,N\phi_{\perp}+Na^3\mathscr{E}\} \ ,\\
0&\approx a^{3\mathscr{W}}\phi_{\perp}+\mathscr{E}_0 \ ,
}
with $f\equiv f(a,p_a;\varphi,p_{\varphi})$ being a phase-space function. As $\mathscr{E}_0$ is a conserved quantity (constant of Nature) in this toy model, we are thus in a similar situation to the relativistic particle case discussed in Sec. \ref{sec:time-conjugate-constant}, with $\mathscr{E}_0$ being analogous to the particle's rest energy. In this way, we can promote $p_{\eta}\Def\Lambda_0\mathscr{E}_0$ to a new, conserved degree of freedom, and we enlarge the set of variables from $(a,p_a,\varphi,p_{\varphi})$ to include the new canonical pair $(\eta,p_{\eta})$ by adding to the action the new term $p_{\eta}\dot{\eta}$. As in the particle case, $\eta$ is a cyclic variable that serves a physical clock that defines time as a canonical conjugate to a constant of Nature. By imposing the time parametrization $\eta(\tau) = \tau$, we fix the lapse function to be $N = \Lambda_0a^{3\mathscr{W}}$, and we can solve the constraint $a^{3\mathscr{W}}\phi_{\perp}+\mathscr{E}_0\approx0$ for $\mathscr{E}_0$ to obtain the on-shell gauge-fixed equations of motion 
\eq{
\dot{f} &= \{f,\Lambda_0a^{3\mathscr{W}}\phi_{\perp}\} \ ,
}
with $\Lambda_0a^{3\mathscr{W}}\phi_{\perp}$ being a conserved quantity. This is identical to the Fock--Stueckelberg results in Eqs. \eref{cosm-eom-constant} and \eref{cosm-eom-constant-ham}, which shows the equivalence of the Fock--Stueckelberg approach with the definition of time as a canonical conjugate to a constant of Nature in this case.\footnote{It is worthwhile to mention that promoting $p_{\eta}\Def\Lambda_0\mathscr{E}_0$ to a new degree of freedom with respect to which the Hamiltonian constraint can be solved corresponds to acknowledging that the energy contribution from homogeneous perfect fluid can be treated not as a mere ``potential term'' dependent on $a$, but rather as a new genuine variable. This is in line with the Schutz formalism for perfect fluids \cite{Schutz}, which is frequently used in the context of canonical cosmology of homogeneous models in order to simplify the solution to the Hamiltonian constraint (see, e.g., \cite{LapRuba,Alvarenga,Almeida,Molinari} and references therein for details and applications). In a nutshell, this formalism allows us to decompose the fluid's four-velocity in terms of velocity potentials, $U_{\mu} = 1/h\left(\partial_{\mu}\epsilon+\zeta\partial_{\mu}\beta+\theta\partial_{\mu}\mathtt{s}\right)$, which are related to the description of the fluid. For example, $h$ is the specific enthalpy, $\mathtt{s}$ is the specific entropy and $\zeta$ and $\beta$ are connected with circulation (vortex motion) and they vanish in isotropic models. With this decomposition and thermodynamical relations, one can establish a series of canonical transformations to establish that the Hamiltonian of the homogeneous perfect fluid can be written as $Np_Ta^{-3\mathscr{W}}$, where $p_T$ is a momentum canonically conjugate to a fluid variable related to its specific entropy and other potentials. This clearly agrees with the Fock--Stueckelberg and constants of Nature approaches if we identify $p_T$ with $p_{\eta}$ and take $\Lambda_0=1$. Thus, one can say that, in this cosmological setting, the Fock--Stueckelberg, constants of Nature and Schutz formalisms are equivalent when it comes to the relaxation of the Hamiltonian constraint.}

The unimodular gravity case with $\mathscr{W}=-1$ is a particular case of the equivalence of the Fock--Stueckelberg and constants of Nature approaches not only in the context of homogeneous cosmology. In the full field-theoretic case, a similar analysis can be carried out for unimodular gravity. We find from Sec. \ref{sec:gug} that the Fock--Stueckelberg field equations with $\mathscr{W}=-1$ read [cf. Eqs. \eref{FS-eom} and \eref{gug-uni} with explicit spatial integrals]
\eq{\label{eq:unimodular-FS-eom-1}
\dot{f} &= \left\{f,\int\D^3y\,\left(\frac{\Lambda_0}{\sqrt{h}}\phi_{\perp}+N^a\phi_a\right)_{(\tau,\mathbf{y})}\right\} \ ,\\
\phi_a &\approx 0 \ , 
}
with $f$ being a phase-space function, and $\Lambda_0\phi_{\perp}/\sqrt{h}$ being a conserved quantity due to Eq. \eref{sec-Ham-const} with $\mathscr{W}\equiv-1$. The same results can be obtained by starting with the parent theory being general relativity with units in which $G=1$ and with a cosmological constant $\mathscr{E}$ (which can be regarded as a constant of Nature, insofar as it is a spacetime constant and one does not consider more complicated dark energy models), and subsequently promoting $\Lambda_0\mathscr{E}/8\pi$ to a conserved field canonically conjugate to a field $\eta(\tau,\mathbf{x})$. As before, we define time via the condition $\eta(\tau,\mathbf{x}) = \tau$, which fixes the lapse to be $N = \Lambda_0/\sqrt{h}$. We can solve the Hamiltonian constraint $\phi_{\perp}+\mathscr{E}\sqrt{h}/8\pi\approx0$ for $\mathscr{E}$, which leads to on-shell partially gauge-fixed field equations that coincide with those of the Fock--Stueckelberg approach given in Eq. \eref{unimodular-FS-eom-1}.

\section{\label{sec:reappear}The reappearance of time in quantum gravity}
The promotion of constants of Nature to spacetime fields and the associated deparametrization of constraints discussed in Sec. \ref{sec:constants}, which yields equivalent results to the Fock--Stueckelberg approach at least in regions where the deparametrization is valid (cf. Sec. \ref{sec:equiv-constant-FS}), leads to a strategy to solve the ``problem of time'' in quantum gravity. This problem is a consequence of spacetime diffeomorphism invariance (which is a local symmetry or a canonical gauge symmetry), and it concerns the lack of a preferred time parameter in general relativity with respect to which the evolution of the wave functional of gravitational and matter fields can be described.

The Dirac quantization of general relativity's constraints leads to timeless equations (cf. Sec. \ref{sec:canonicalGR})\footnote{The quantum constraints are timeless in the sense that no explicit time parameter appears, although of course one may attempt to interpret one of the physical degrees of freedom as an ``internal'' time \cite{Kiefer:book}.}
\eq{
\hat{\phi}_{\perp}\!\ket{\Psi} = 0 \ , \ \hat{\phi}_a\!\ket{\Psi} = 0 \ ,
}
which imply a timeless constraint $\hat{H}\ket{\Psi} = \lambda^a(\tau)\hat{\phi}_a\ket{\Psi} = 0$ (spatial integration and summation over discrete indices implied). Due to the structure of $\phi_{\perp}$ (cf. Sec. \ref{sec:canonicalGR}), the timeless quantum constraints lead to a generalization of the Klein--Gordon equation for the relativistic particle, and thus the problem of time in quantum gravity is inevitably linked to difficulties with the definition of the Hilbert space.\footnote{See, however, \cite{Marolf,Giulini} for a promising way to define a positive-definite inner product at least for symmetry-reduced toy models.} In fact, the Hamiltonian constraint of canonical quantum cosmology of homogeneous models coincides with that obtained from mechanical models discussed in Sec. \ref{sec:mech}, which is nothing but a Klein--Gordon equation for a relativistic particle with a general potential and a general background (configuration-space) metric.

In contrast, the relaxation of the (Hamiltonian) constraint obtained with the Fock--Stueckelberg or constants of Nature approaches leads to the ordinary time-dependent (functional) Schr\"odinger equation
\eq{
\I\hbar\del{}{\tau}\ket{\Psi} = \hat{H}\ket{\Psi} \ ,
}
due to the fact that the time translations are no longer gauged. One can thus speak of the ``reappearance of time,'' which is nothing but the breaking of spacetime diffeomorphism invariance engendered by the Fock--Stueckelberg construction. It is important to emphasize again that this construction is not related to the consistency of the quantization procedure itself, but rather to the definition of a different \emph{classical} theory to be subsequently quantized. It makes the quantization more conceptually and technically straightforward at the cost of the theory's local symmetry, and it was one of the motivations for Stueckelberg to develop his approach for the relativistic particle (by breaking the worldline diffeomorphism invariance), as well as for modern researchers. Indeed, from the ``original'' unimodular gravity to its generalizations \cite{HT-CC-1,UnruhUni,UnruhWald,HT-CC-2,Kaplan1,we,Barv,Barv1,Kam-Tron-Ven,Kara}, and with several proposals to define time as the canonical conjugate to different constants of Nature \cite{CG1,CG2,CG3,GT1,GT2,GT3,GT4,Magueijo1,Magueijo2,Magueijo3,Magueijo4,Magueijo5,Magueijo6,Magueijo7,Kaplan}, this strategy has led to a series of articles claiming to solve the problem of time in quantum gravity.

It is worthwhile to note, however, that this solution to the problem of time entails the introduction of new physical fields, one of which plays the role of a (possibly global) physical clock. Whether such a field independently exists in Nature is an open question, and otherwise the formalism risks being a mere effective treatment of different physical clocks, each of which is valid only in certain regions, in which case the simplification brought by the deparametrization relative to constants of Nature seems vacuous -- the same difficulties appear with the traditional gauge fixing of the $z$ fields; i.e. the traditional problem of multiple choices of clocks. If no such independent, new physical clock field can be measured, this solution to the problem of time is at risk of being no solution at all. Instead of achieving a consistent and physically meaningful implementation of time reparametrizations in the quantum theory of general models, the Fock--Stueckelberg and constants of Nature approaches simply break the symmetry\footnote{The symmetry is broken in the sense that the Fock--Stueckelberg action in Eq. \eref{FS-action-1} or the action in Eq. \eref{depar-action} that appears in the constants of Nature approach display a reduced (or residual) set of symmetries with respect to the original action in Eq. \eref{action}.} that causes the problem in the first place, and thus they may be a way of avoiding rather than solving the problem.

\section{\label{sec:conclusions}Conclusions}
Generally, one can say that in a theory with gauge symmetries or in a theory with first-class constraints, the relaxation of some of these constraints implies the appearance of new physical degrees of freedom. This is the essence of the Fock--Stueckelberg approach that we have discussed in this article. The idea is rather simple, and it is often independently rediscovered in the literature, whereas its origins and physical meaning frequently remain unclear. Our goal was to present a comprehensive overview of a method that generalizes the original works of Fock and Stueckelberg, which we refer to as the Fock--Stueckelberg theory.

The physical meaning of the Fock--Stueckelberg theory was best captured by Dirac in his own proposal of a new electrodynamics \cite{Dirac}: if we do not know the structure and dynamics of the ``sources'' of gauge fields (such as the electrons in electromagnetism or certain kinds of matter-energy in gravity), we can consider breaking the gauge symmetry, and promoting some of the ``spurious" gauge degrees of freedom to physical fields that describe or define the sources themselves. So are Dirac's ``electrons without electrons,'' and the ``matter without matter'' of generalized unimodular gravity, which can lead to models of dark matter and dark energy without details about their internal structure. So is the independent mass ``field'', which leads to an independent ``proper time'' variable, in Fock's and Stueckelberg's relativistic particles, which do not suffer from the traditional problem of time of Klein--Gordon-like Hamiltonian constraints, and thus they inspire avoidances of this problem in quantum gravity.

In the literature, the many methods that are related to the Fock--Stueckelberg approach may thus lead to useful models of the sources of the relaxed gauge theories. In this regard, it is worthwhile to mention some recent works that have appeared after the submission of the present article. In \cite{DelGrosso}, the simultaneous relaxation of the constraints of general relativity and of electromagnetism (a Dirac-type electrodynamics) produces a very particular kind of electric charges. Moreover, it also possible that new developments may lead to extensions of the Fock--Stueckelberg mechanism. In \cite{MagueijoNovo,MagueijoNovo2}, preferred types of foliations (in particular, geodesic foliations) are considered in a type of relaxation of general relativity. There, it is shown that a kind of ``dark dust'' and more general types of matter may emerge. Interestingly, one seems to obtain results that are different from the applications of generalized unimodular gravity for effective matter fields other than ``dark dust.'' This is because the effective fluid has anisotropic stress under certain conditions.\footnote{We thank Jo\~{a}o Magueijo for bringing these results to our attention.} The particular case of ``dark dust'' had also been considered in the older reference \cite{Mukohyama} (see also \cite{Kobakhidze}), without, however, a discussion of the connection with (generalized) unimodular gravity or the Fock--Stueckelberg gauge fixation, but including the discussion of another formalism for relaxation of the Hamiltonian constraint: Ho\v{r}ava--Lifschitz gravity. It would be interesting to ascertain whether the recent results of \cite{MagueijoNovo,MagueijoNovo2} are equivalent to a Fock--Stueckelberg gauge fixation or to an extension of it.

Although other approaches can consider a breaking of gauge symmetries, the Fock--Stueckelberg theory does so in an economical way: it promotes a gauge-fixing condition to a physical equation on a par with the first-class constraints in the action. This breaks the symmetry by relaxing the original constraints $\phi_a$, but if we interpret the new degrees of freedom $w$ as a new form of matter, we can consider that new constraints arise, $\phi_a+w\approx0$ [cf. Eq. \eref{FS-constraints}]. The consistency of the new Fock--Stueckelberg constraints has to be analyzed. In certain cases, the new degrees of freedom can be identified with constants of Nature promoted to physical fields (cf. Sec. \ref{sec:constants}), and the new constraints $\phi_a+w\approx0$ might still generate the original symmetry, although they are to be deparametrized with respect to the $w$ fields or the constants of Nature, thereby freeing the restriction on the values of $\phi_a$.

It is clear that the Fock--Stueckelberg approach generates a classical theory, possibly different from its parent theory. The Fock--Stueckelberg theory can be subsequently quantized with the usual methods, from Dirac to BRST quantization \cite{Hanson,HT:book}. This goes against some recent claims made in \cite{Kaplan,Kaplan1,Kaplan2} that fixing the gauge \`{a} la Fock and Stueckelberg (i.e., fixing the gauge at the level of the action before variation) is necessary in order to make the quantum theory or, more precisely, the quantum Hamiltonian and the path integral of gauge theories well defined. Those can be well defined, presumably, via the usual methods, provided field-theoretic divergences (which come from functional derivatives, for example) can be regularized, an issue about which the Fock--Stueckelberg approach remains silent. Thus, the physical meaning of this formalism need not be tied to the consistency of the quantization of gauge theories, and the fundamental quantum theory need not break gauge invariance. However, it is noteworthy that, since the Fock--Stueckelberg mechanism often leads to a more straightforward quantization, it is often motivated with that end in mind since the original works of Fock and Stueckelberg \cite{Fock,Stueck,Stueck1,Stueck2}.

So what is the value of the Fock--Stueckelberg formalism? Its value lies in its ability to simplify a local symmetry to a residual gauge invariance by defining a kind of ``shadow'' matter (such as Dirac's electrons) from the original ``spurious'' gauge degrees of freedom. It frees the theorist, as it did Dirac, to explore new phenomenology without introducing many new complications, and, as discussed above, it may even lead to a solution to the problem of time that is somewhat trivial. Although advantageous, this economy comes at the cost of postulating new fields, and eventually it might become necessary to ascertain whether such degrees of freedom really exist and what their fundamental structure is. Going forward, we believe that the Fock--Stueckelberg method might be useful as an effective approach to deal with hitherto poorly understood problems in phenomenology (such as dark matter and dark energy), but a more fundamental theory may require us to move beyond this simple idea of Fock and Stueckelberg. Hopefully this article can serve as a reference to help us move closer to a fundamental understanding of classical and quantum gauge theories.

\acknowledgments{The authors thank the anonymous referees for their constructive feedback. The work of R.C., A.K. and A.T. was partially supported by the INFN Grant FLAG. The work by F.G.P. was partially supported by the INFN Grant  STEFI. The work of L.C. is supported by the Basque Government Grant \mbox{IT1628-22}, and by the Grant PID2021-123226NB-I00 (funded by MCIN/AEI/10.13039/501100011033 and by “ERDF A way of making Europe”). It is also partly funded by the IKUR 2030 Strategy of the Basque Government.}

\begin{appendix}
\section{\label{app:noether}Noether charge}
In this appendix, we recall the derivation of the equation for the Noether charge. The reader is referred to \cite{HT:book,Pons,Sundermeyer,Chataig:Thesis,Hanson,SalisburyRosenfeld} for further details.

In $d+1$ spacetime dimensions, let us consider the Lagrangian
\begin{equation}\begin{aligned}\label{eq:noether-lagrangian}
L = \int\mathrm{d}^dx\, \mathcal{L}(\xi(\tau,\mathbf{x}),\partial_{\mu}{\xi}(\tau,\mathbf{x})) \ ,
\end{aligned}\end{equation}
for the fields $\xi^{i}(\tau,\mathbf{x})$ ($i = 1,\ldots, X$), with $(\tau,\mathbf{x})$ comprising the spacetime point $x^{\mu}$ and $\partial_{\mu}\xi\equiv \partial \xi/\partial x^{\mu}$. We consider field derivatives only up to first order for simplicity. In the mechanical case, $d = 0$ and there is no space nor spatial derivatives nor spatial integration. Now consider the one-parameter family of transformations
\begin{equation}\begin{aligned}\label{eq:noether-transf}
x^{\mu}&\mapsto x^{\mu}(s) \ , \\
\xi(\tau,\mathbf{x}) &\mapsto \xi(\tau(s),\mathbf{x}(s);s) \ ,
\end{aligned}\end{equation}
which are assumed to be continuously connected to the identity, and they may include not only changes in the $\xi$ fields themselves but also in the spacetime point (with $x^{\mu}(0) = x^{\mu}$). We have
\begin{equation}\begin{aligned}
\frac{\mathrm{d}}{\mathrm{d}s}\xi(\tau(s),\mathbf{x}(s);s) = \frac{\partial \xi}{\partial s}+\frac{\partial x^{\mu}}{\partial s}\frac{\partial \xi}{\partial x^{\mu}} \ , 
\end{aligned}\end{equation}
where we include a tacit summation over repeated indices, as usual, and we note that $[\partial_s,\partial_{\mu}] = 0$. The partial derivative $\partial_s$ is related to an ``active variation'' of the fields $\xi$ under the transformation given in Eq.~\eref{noether-transf}, i.e., to a change in the fields with a fixed spacetime point.\footnote{This terminology recalls, for example, the active rotation of a vector, in which the vector is changed but the coordinate system is kept fixed.} 

As a matter of notational convenience, we define the operator
\begin{equation}\begin{aligned}\label{eq:delta-op}
\delta := s\left.\frac{\partial}{\partial s}\right|_{s = 0}\!\equiv s\left(\frac{\mathrm{d}}{\mathrm{d}s}-\frac{\partial x^{\mu}}{\partial s}\frac{\partial}{\partial x^{\mu}}\right)_{s = 0}  \ , 
\end{aligned}\end{equation}
which can be defined for any $s$, but it also yields the active variation of $\xi$ if $s$ is, in particular, very small (``infinitesimal''):
\begin{equation}\begin{aligned}
\xi(\tau(0),\mathbf{x}(0);s) - \xi(\tau,\mathbf{x}) = \delta \xi + \mathcal{O}(s^2)\ .
\end{aligned}\end{equation}
Using the Jacobian
\begin{equation}\begin{aligned}\label{eq:noether-jacobian}
\det\left[\frac{\partial x^{\mu}(s)}{\partial x^{\nu}}\right] = 1+s\frac{\partial}{\partial x^{\mu}}\left.\frac{\mathrm{d}}{\mathrm{d}s}\right|_{s = 0}x^{\mu}(s)+\mathcal{O}(s^2) \ ,
\end{aligned}\end{equation}
and assuming that the term linear in $s$ is not identically zero, we can now compute the variation of the action to be
\begin{equation}\begin{aligned}
&S[\xi(\tau(s),\mathbf{x}(s);s)]-S[\xi(\tau,\mathbf{x})] \\
&= \int\mathrm{d}^{d+1}x\, \left\{\left[\frac{\partial\mathcal{L}}{\partial \xi^i}-\frac{\partial}{\partial x^{\mu}}\frac{\partial\mathcal{L}}{\partial(\partial_{\mu}\xi^i)}\right]\delta \xi^i\right.\\
&\left.\ \ \ +\frac{\partial}{\partial x^{\mu}}\left[\frac{\partial\mathcal{L}}{\partial(\partial_{\mu}\xi^i)}\delta\xi^i+s\mathcal{L}\left.\frac{\mathrm{d}}{\mathrm{d}s}\right|_{s = 0}\!\!x^{\mu}(s)\right]\right\} +\mathcal{O}(s^2)\ , 
\end{aligned}\end{equation}
where $\delta \xi^i$ is defined according to Eq.~\eref{delta-op}. Only this ``active variation'' of the fields appears because the change in the fields due to the change in spacetime coordinates is canceled by the change in the integration measure [the contribution from the Jacobian factor in Eq.~\eref{noether-jacobian}]. The transformation is then a symmetry if it does not alter the form of the field equations, which means that the Lagrangian must change by a total time derivative:
\begin{equation}\begin{aligned}\label{eq:pre-noether}
&\int\mathrm{d}^{d}x\, \left\{\left[\frac{\partial\mathcal{L}}{\partial \xi^i}-\frac{\partial}{\partial x^{\mu}}\frac{\partial\mathcal{L}}{\partial(\partial_{\mu}\xi^i)}\right]\delta \xi^i\right.\\
&\left. \ \ \ +\frac{\partial}{\partial x^{\mu}}\left[\frac{\partial\mathcal{L}}{\partial(\partial_{\mu}\xi^i)}\delta\xi^i+s\mathcal{L}\left.\frac{\mathrm{d}}{\mathrm{d}s}\right|_{s = 0}\!\!x^{\mu}(s)\right]\right\} = \frac{\mathrm{d}F}{\mathrm{d}\tau} \ .
\end{aligned}\end{equation}
We now assume that $\int\mathrm{d}^dx\,\partial_{\mathbf{x}}(\ldots) = 0$ even off-shell (i.e., we restrict ourselves to field configurations that satisfy this property by means of appropriate fall-off conditions, even if they do not satisfy the field equations). Then, instead of working with $F$, we define
\begin{equation}\begin{aligned}
G =  \int\mathrm{d}^{d}x\,\left[\frac{\partial\mathcal{L}}{\partial(\partial_{\tau}\xi^i)}\delta\xi^i+s\mathcal{L}\left.\frac{\mathrm{d}}{\mathrm{d}s}\right|_{s = 0}\!\!\tau(s)\right]-F \ , 
\end{aligned}\end{equation}
so that we can conveniently rewrite Eq.~\eref{pre-noether} as
\begin{equation}\begin{aligned}\label{eq:noether}
 \int\mathrm{d}^{d}x\, \left[\frac{\partial\mathcal{L}}{\partial \xi^i}-\frac{\partial}{\partial x^{\mu}}\frac{\partial\mathcal{L}}{\partial(\partial_{\mu}\xi^i)}\right]\delta \xi^i+\frac{\mathrm{d}G}{\mathrm{d}\tau} = 0 \ .
\end{aligned}\end{equation}
This general result leads to both Noether theorems (see, e.g., \cite{HT:book,Pons,Sundermeyer,Chataig:Thesis,Hanson,SalisburyRosenfeld}). Indeed,  $G$ is the Noether charge, which is conserved on-shell (i.e., if the field equations are satisfied). Evidently, the precise form of $G$ depends on the Lagrangian and on the symmetry transformation considered, and it can be determined from the off-shell expression in Eq.~\eref{noether} (which is valid even if the field equations are not satisfied).

As a simple example, consider the mechanical case ($d = 0$) with $\xi = (q,p)$ and Lagrangian
\begin{equation}\begin{aligned}
L = p\dot{q}-\frac{p^2}{2m}-V(q) \ , 
\end{aligned}\end{equation}
as well as a global active time translation $\tau\mapsto\tau$, $\xi(\tau)\mapsto \xi(\tau;s) = \xi(\tau+s)$. The above results lead to $\delta \xi = s\dot{\xi}(\tau)$ which, together with Eq.~\eref{noether}, yields
\begin{equation}\begin{aligned}
G = s\left[\frac{p^2}{2m}+V(q)\right]+\text{constant}\,.
\end{aligned}\end{equation}
Since $s$ does not depend on $\tau$, the on-shell conservation of $G$ implies the on-shell conservation of the Hamiltonian (regardless of whether $s$ is very small), which is the elementary result for global time translation invariance. Evidently, we can define $G$ without any $s$ dependence by repeating the derivation of Eq.~\eref{noether} using only $\partial/\partial s|_{s = 0}$ instead of the $\delta$ defined in Eq.~\eref{delta-op}. This yields a charge $G$ that coincides with the Hamiltonian (up to an additive constant-in-time factor) in the case of global time translations.

For local (gauge) symmetries, the functional form of $G$ may be ascertained from the off-shell equation~\eref{noether}. If we adopt DeWitt's compact notation (cf. Sec.~\ref{sec:gauge-transf}), and we take the fields to be $\xi = (z,\lambda)$ with the action given in Eq.~\eref{action} [assuming that both the symplectic potential and the Hamiltonian may depend on $z$ and its first-order spatial derivatives for consistency with Eq.~\eref{noether-lagrangian}], then Eq.~\eref{noether} becomes Eq.~\eref{Noether}, from which we can deduce the form of $G$ and its role as the symmetry generator via the Poisson bracket [cf. Eq. \eref{Noether2}].
\end{appendix}


\begin{thebibliography}{99}
\bibitem{Rosenfeld}
L. Rosenfeld, {\it Zur Quantelung der Wellenfelder}, \href{https://doi.org/10.1002/andp.19303970107}{Ann. Phys. (Berl.) {\bf 397}, 113 (1930)}.

\bibitem{SalisburyRosenfeld}
D.~Salisbury and K.~Sundermeyer,
{\it L\'eon Rosenfeld\textquoteright{}s general theory of constrained Hamiltonian dynamics},
\href{https://doi.org/10.1140/epjh/e2016-70042-7}{Eur. Phys. J. H \textbf{42}, 23 (2017)}.

\bibitem{Dirac1}
P. A. M. Dirac, {\it Generalized Hamiltonian Dynamics}, \href{https://doi.org/10.4153/CJM-1950-012-1}{Can. J. Math. {\bf2}, 129 (1950)}.

\bibitem{Dirac2}
P. A. M. Dirac, {\it Generalized Hamiltonian dynamics}, \href{https://doi.org/10.1098/rspa.1958.0141}{Proc. R. Soc. A {\bf 246}, 326 (1958)}.

\bibitem{Dirac3}
P. A. M. Dirac, {\it Lectures on Quantum Mechanics}, Dover Publications, Inc., New York (2001). Originally published: Belfer Graduate School of Science, Yeshiva University, New York (1964).

\bibitem{Bergmann}
P. G. Bergmann,
{\it Conservation Laws in General Relativity as the Generators of Coordinate Transformations},
\href{https://doi.org/10.1103/PhysRev.112.287}{Phys. Rev. {\bf 112}, 287 (1958)};
{\it Observables in General Relativity},
\href{https://doi.org/10.1103/RevModPhys.33.510}{Rev. Mod. Phys. {\bf 33}, 510 (1961)};
{\it ``Gauge-Invariant'' Variables in General Relativity},
\href{https://doi.org/10.1103/PhysRev.124.274}{Phys. Rev. {\bf 124}, 274 (1961)}.

\bibitem{RBD}
D.~C.~Salisbury,
{\it Rosenfeld, Bergmann, Dirac and the invention of constrained Hamiltonian dynamics},
\href{https://doi.org/10.1142/9789812834300\_0435}{The Eleventh Marcel Grossmann Meeting, 2467 (2008)}.

\bibitem{Hanson}
A.~J.~Hanson, T.~Regge and C.~Teitelboim,
{\it Constrained Hamiltonian Systems},
Accademia Nazionale dei Lincei (1976).

\bibitem{HT:book}
M. Henneaux and C. Teitelboim, {\it Quantization of Gauge Systems}, Princeton University Press, Princeton, NJ (1992).

\bibitem{DiracGravity}
P. A. M. Dirac, {\it The theory of gravitation in Hamiltonian form}, \href{https://doi.org/10.1098/rspa.1958.0142}{Proc. R. Soc. A {\bf246}, 333 (1958)}.

\bibitem{Kiefer:book}
C. Kiefer, {\it Quantum Gravity}, 3rd ed., International Series of Monographs on Physics, Oxford University Press, Oxford (2012).

\bibitem{Anderson:book}
E. Anderson, {\it The Problem of Time -- Quantum Mechanics Versus General Relativity}, in: \href{https://doi.org/10.1007/978-3-319-58848-3}{Fundamental Theories of Physics Vol 190}, Springer International Publishing, Cham (2017).

\bibitem{Kaplan}
A. K. Burns, D. E. Kaplan, T. Melia and S. Rajendran, {\it Time Evolution in Quantum Cosmology}, 
\href{https://doi.org/10.48550/arXiv.2204.03043}{arXiv:2204.03043 [gr-qc]}.

\bibitem{HT-CC-1}
M.~Henneaux and C.~Teitelboim, {\it The Cosmological Constant as a Canonical Variable},
\href{https://doi.org/10.1016/0370-2693(84)91493-X}{Phys. Lett. B \textbf{143}, 415 (1984)}.

\bibitem{UnruhUni}
W. G. Unruh,  {\it A unimodular theory of canonical quantum gravity}, \href{https://doi.org/10.1103/PhysRevD.40.1048}{Phys. Rev. D {\bf 40}, 1048 (1989)}.
    
\bibitem{UnruhWald}
W.~G.~Unruh and R.~M.~Wald,
{\it Time and the Interpretation of Canonical Quantum Gravity},
\href{https://doi.org/10.1103/PhysRevD.40.2598 }{Phys. Rev. D \textbf{40}, 2598 (1989)}.

\bibitem{HT-CC-2}
M. Henneaux and C. Teitelboim, {\it The cosmological constant and general covariance}, \href{https://doi.org/10.1016/0370-2693(89)91251-3}{Phys. Lett. B {\bf 222}, 195 (1989)}.

\bibitem{CG1}
A.~Carlini and J.~Greensite, {\it Fundamental constants and the problem of time},
\href{https://doi.org/10.1103/PhysRevD.52.936}{Phys. Rev. D \textbf{52}, 936 (1995)}.

\bibitem{CG2}
A.~Carlini and J.~Greensite,
{\it Square root actions, metric signature, and the path integral of quantum gravity},
\href{https://doi.org/10.1103/PhysRevD.52.6947}{Phys. Rev. D \textbf{52}, 6947 (1995)}.

\bibitem{CG3}
A.~Carlini and J.~Greensite,
{\it The Mass shell of the universe},
\href{https://doi.org/10.1103/PhysRevD.55.3514}{Phys. Rev. D \textbf{55}, 3514 (1997)}.

\bibitem{GT1}
S.~Gryb and K.~Th\'ebault,
{\it Schrodinger Evolution for the Universe: Reparametrization},
\href{https://doi.org/10.1088/0264-9381/33/6/065004}{Class. Quant. Grav. \textbf{33}, 065004 (2016)}.

\bibitem{GT2}
S.~Gryb and K.~P.~Y.~Th\'ebault,
{\it Bouncing Unitary Cosmology I: Mini-Superspace General Solution},
\href{https://doi.org/10.1088/1361-6382/aaf823}{Class. Quant. Grav. \textbf{36}, 035009 (2019)};
{\it Bouncing Unitary Cosmology II: Mini-Superspace Phenomenology,}
\href{https://doi.org/10.1088/1361-6382/aaf837}{Class. Quant. Grav. \textbf{36}, 035010 (2019)}.

\bibitem{GT3}
S.~Gryb and K.~P.~Y.~Th\'ebault,
{\it Superpositions of the cosmological constant allow for singularity resolution and unitary evolution in quantum cosmology},
\href{https://doi.org/10.1016/j.physletb.2018.08.013}{Phys. Lett. B \textbf{784}, 324 (2018)}.

\bibitem{GT4}
S.~Gryb and K.~P.~Y.~Th\'ebault,
{\it Time Regained: Symmetry and Evolution
in Classical Mechanics}, Oxford University Press, Oxford (2024).

\bibitem{Magueijo1}
S.~Alexander, J.~Magueijo and L.~Smolin,
{\it The Quantum Cosmological Constant},
\href{https://doi.org/10.3390/sym11091130}{Symmetry \textbf{11}, 1130 (2019)}.

\bibitem{Magueijo2}
J.~Magueijo and L.~Smolin,
{\it A Universe that does not know the time},
\href{https://doi.org/10.3390/universe5030084}{Universe \textbf{5}, 84 (2019)}.

\bibitem{Magueijo3}
J.~Magueijo,
{\it Cosmological time and the constants of nature},
\href{https://doi.org/10.1016/j.physletb.2021.136487}{Phys. Lett. B \textbf{820}, 136487 (2021)}.

\bibitem{Fock}
V. Fock, {\it Sobstvennoe vremya v klassichskoj i kvantovoj mekhanike} (in Russian), Izv. AN SSSR, Ser. Fizika, No 4-5, 551 (1937); republished in the book {\it Raboty po kvantovoj teorii polya (Works on Quantum Field Theory)}, publishing house of Leningrad University (1957); also published as {\it Die Eigenzeit in der klassischen und in der Quantennechanik}, Phys. Zeit. d. Sowjetunion {\bf12}, 404 (1937).

\bibitem{Stueck}
E. C. G. Stueckelberg, {\it Remarque \`a propos de la cr\'eation de paires de particules en th\'eorie de relativit\'e},  Helv. Phys.  Acta {\bf 14}, 588  (1941).

\bibitem{Stueck1}
E. C. G. Stueckelberg, {\it La signification du temps propre en m\'ecanique ondulatoire}, Helv. Phys.  Acta {\bf 14}, 322  (1941).

\bibitem{Stueck2}
E. C. G. Stueckelberg, {\it La m\'ecanique du point mat\'eriel en th\'eorie de
relativit\'e et en th\'eorie des quants}, Helv. Phys.  Acta {\bf 15}, 23  (1942).

\bibitem{Chataig:Thesis}
L. Chataignier, {{\it Timeless Quantum Mechanics and the Early Universe}},  \href{https://doi.org/10.1007/978-3-030-94448-3}{Springer Theses} (Springer, Cham, Switzerland, 2022).

\bibitem{Lanczos}
C. Lanczos, {\it The variational principles of mechanics}, 4th edn., Dover, New York (1970).

\bibitem{GrybJ}
S.~B.~Gryb,
{\it Jacobi's principle and the disappearance of time},
\href{https://doi.org/10.1103/PhysRevD.81.044035}{Phys. Rev. D \textbf{81}, 044035 (2010)}.

\bibitem{HPStueck1}
L. P. Horwitz and C. Piron, {\it Relativistic dynamics}, Helv. Phys. Acta {\bf46}, 316 (1973).

\bibitem{HPStueck2}
J. R. Fanchi and R. E. Collins, {\it Quantum mechanics of relativistic spinless particles}, \href{https://doi.org/10.1007/BF00715059}{Found. Phys. {\bf8}, 851 (1978)}.

\bibitem{HPStueck3}
L. P. Horwitz, {\it Relativistic Quantum Mechanics}, Fundamental Theories of Physics (Springer, Dordrecht, NL, 2015).

\bibitem{Kaplan1}
D. E. Kaplan, T. Melia and S. Rajendran,
{\it The Classical Equations of Motion of Quantized Gauge Theories, Part I: General Relativity}, \href{https://doi.org/10.48550/arXiv.2305.01798}{arXiv:2305.01798 [hep-th]}.

\bibitem{Kaplan2}
D. E. Kaplan, T. Melia and S. Rajendran, 
{\it The Classical Equations of Motion of Quantized Gauge Theories, Part 2: Electromagnetism},
\href{https://doi.org/10.48550/arXiv.2307.09475}{arXiv:2307.09475 [hep-th]}. 

\bibitem{Magueijo4}
J.~Magueijo,
{\it Evolving laws and cosmological energy},
\href{https://doi.org/10.1103/PhysRevD.108.103514}{Phys. Rev. D \textbf{108}, 103514 (2023)}.

\bibitem{Magueijo5}
P.~M.~Bassani and J.~Magueijo,
{\it Unimodular-like times, evolution and Brans-Dicke Gravity},
\href{https://doi.org/10.1142/S0218271823501134}{Int. J. Mod. Phys. D \textbf{33}, 2350113 (2024)}.

\bibitem{Magueijo6}
A.~Etkin, J.~Magueijo and F.~S.~Rassouli,
{\it Vortices, topology and time},
\href{https://doi.org/10.1016/j.physletb.2024.138810}{Phys. Lett. B \textbf{855}, 138810 (2024)}.

\bibitem{Einstein}
A. Einstein, Sitzungsber. Preuss. Akad. Wiss. Berlin (Math. Phys.) {\bf 1919}, 349 (1919). 

\bibitem{Arnold}
V. Arnold,
{\it Sur la g\'eom\'etrie differentielle des groupes de Lie de dimension infinie et ses application \`a l'hydrodynamique des fluides parfaits}, Ann. De l'Institut Fourier {\bf 16}, 319 (1966).

\bibitem{Wolski}
A. Wolski and J. S. Dowker,
{\it Area-preserving diffeomorphisms of Riemann surfaces},
\href{https://doi.org/10.1063/1.529153}{J. Math. Phys. {\bf 32}, 2304 (1991)}.

\bibitem{Bose}
S.K. Bose and S.A. Bruce,
{\it Classical symmetries of a closed bosonic 3-brane},
\href{https://doi.org/10.1016/0370-2693(89)90577-7}{Phys. Lett. B {\bf 225}, 331 (1989)};
{\it Algebras for the two-sphere and the three-sphere groups of compact simple Lie groups},
\href{https://doi.org/10.1063/1.528835}{J. Math. Phys. {\bf 31}, 2346 (1990)}.

\bibitem{Arakelian}
T. A. Arakelian and G. K. Savvidy,
{\it Cocycles of Area Preserving Diffeomorphisms and Anomalies in Theory of Relativistic Surfaces},
\href{https://doi.org/10.1016/0370-2693(88)91375-5}{Phys. Lett. B {\bf 214}, 350 (1988)};
{\it Geometry of a group of area-preserving diffeomorphisms},
\href{https://doi.org/10.1016/0370-2693(89)90916-7}{Phys. Lett. B {\bf 223}, 41 (1989)}.

\bibitem{Floratos}
E.G. Floratos and J. Iliopoulos,
{\it A Note on the Classical Symmetries of the Closed Bosonic Membranes},
\href{https://doi.org/10.1016/0370-2693(88)90220-1}{Phys. Lett. B {\bf 201}, 237 (1988)}.

\bibitem{Bars}
I. Bars, C. N. Pope and E. Sezgin,
{\it Central Extensions of Area Preserving Membrane Algebras},
\href{https://doi.org/10.1016/0370-2693(88)90354-1}{Phys. Lett. B {\bf 210}, 85 (1988)}.

\bibitem{Wit}
B. de Wit, U. Marquard and H. Nicolai,
{\it Area-preserving diffeomorphisms and supermembrane Lorentz invariance},
\href{https://doi.org/10.1007/BF02097044}{Commun. Math. Phys. {\bf 128}, 3 (1990)}.

\bibitem{closed}
A. Yu. Kamenshchik and S. L. Lyakhovich,
{\it Hamiltonian BFV–BRST theory of closed quantum cosmological models},
\href{https://doi.org/10.1016/S0550-3213(97)00203-4}{Nucl. Phys. B {\bf 495}, 309 (1997)}.

\bibitem{Kugo}
T.~Kugo, R.~Nakayama and N.~Ohta,
{\it BRST quantization of general relativity in unimodular gauge and unimodular gravity},
\href{https://doi.org/10.1103/PhysRevD.104.126021}{Phys. Rev. D \textbf{104}, 126021 (2021)}.

\bibitem{Dirac}
P.A.M. Dirac, {\it A new classical theory of electrons}, 
\href{https://doi.org/10.1098/rspa.1951.0204}{Proc. Roy. Soc. London A  {\bf 209}, 291 (1951)}.

\bibitem{post}
R.~Righi and G.~Venturi,
{\it Nonlinear Approach To Electrodynamics},
\href{https://doi.org/10.1007/BF01880265}{Int.\ J.\ Theor.\ Phys.\  {\bf 21}, 63 (1982)}.
  
\bibitem{post1}
R.~Righi and G.~Venturi,
{\it Is The Electric Charge Of Topological Origin?},
\href{https://doi.org/10.1007/BF02776212}{Lett.\ Nuovo Cim.\  {\bf 31}, 487 (1981)}.

\bibitem{post11}
R.~Righi, G.~Venturi, and V.~Zamiralov,
{\it A Non-Abelian Gauge Theory in a Nonlinear Gauge},
\href{https://doi.org/10.1007/BF02896240}{Nuovo Cim. A \textbf{47}, 518 (1978)}.
  
\bibitem{post2}
A.~Akhmeteli,
{\it One real function instead of the Dirac spinor function},
\href{https://doi.org/10.1063/1.3624336}{J.\ Math.\ Phys.\  {\bf 52}, 082303 (2011)}.

\bibitem{post3}
A.~Akhmeteli,
{\it No Drama Quantum Electrodynamics?},
\href{https://doi.org/10.1140/epjc/s10052-013-2371-4}{Eur.\ Phys.\ J.\ C {\bf 73}, no. 4, 2371 (2013)}. 

\bibitem{Vergara}
M.~Henneaux, C.~Teitelboim and J.~D.~Vergara,
Gauge invariance for generally covariant systems,
\href{https://doi.org/10.1016/0550-3213(92)90166-9}{Nucl. Phys. B \textbf{387}, 391 (1992)}. 

\bibitem{DeWittFields}
B.S. DeWitt,
{\it Dynamical Theory of Groups and Fields}, Gordon and Breach, New York (1965).

\bibitem{Pons}
J.~M.~Pons, D.~C.~Salisbury and L.~C.~Shepley,
{\it Gauge transformations in the Lagrangian and Hamiltonian formalisms of generally covariant theories},
\href{https://doi.org/10.1103/PhysRevD.55.658}{Phys. Rev. D \textbf{55}, 658 (1997)};
{\it Gauge transformations in Einstein-Yang-Mills theories},
\href{https://doi.org/10.1063/1.533425}{J. Math. Phys. \textbf{41}, 5557 (2000)}.

\bibitem{FP}
L. D. Faddeev and V. N. Popov,
{\it Feynman diagrams for the Yang-Mills field},
\href{https://doi.org/10.1016/0370-2693(67)90067-6}{Phys. Lett. B {\bf 25}, 29 (1967)};
L. D. Faddeev,
{\it Feyman Integrals for Singular Lagrangians},
Theor. Math. Phys. {\bf1}, 1 (1969).

\bibitem{Magueijo7}
J.~Magueijo,
{\it Mach's principle and dark matter},
\href{https://arxiv.org/abs/2312.07597}{arXiv:2312.07597 [hep-th]}.

\bibitem{varying1}
J.~Magueijo,
{\it New varying speed of light theories},
\href{https://doi.org/10.1088/0034-4885/66/11/R04}{Rept. Prog. Phys. \textbf{66}, 2025 (2003)}.

\bibitem{varying2}
D.~Kimberly and J.~Magueijo,
{\it Varying alpha and the electroweak model},
\href{https://doi.org/10.1016/j.physletb.2004.01.050}{Phys. Lett. B \textbf{584}, 8 (2004)}.

\bibitem{varying3}
G.~Calcagni, J.~Magueijo and D.~Rodr\'\i{}guez Fern\'andez,
{\it Varying electric charge in multiscale spacetimes},
\href{https://doi.org/10.1103/PhysRevD.89.024021}{Phys. Rev. D \textbf{89}, 024021 (2014)}.

\bibitem{varying4}
S.~Alexander, M.~Cort\^es, A.~R.~Liddle, J.~Magueijo, R.~Sims and L.~Smolin,
{\it Zero-parameter extension of general relativity with a varying cosmological constant},
\href{https://doi.org/10.1103/PhysRevD.100.083506}{Phys. Rev. D \textbf{100}, 083506 (2019)};
{\it Cosmology of minimal varying Lambda theories},
\href{https://doi.org/10.1103/PhysRevD.100.083507}{Phys. Rev. D \textbf{100}, 083507 (2019)}.

\bibitem{varying5}
S.~Alexander, T.~Daniel and J.~Magueijo,
{\it The Ashtekar Variables and a Varying Cosmological Constant from Dynamical Chern-Simons Gravity},
\href{https://arxiv.org/abs/2207.08885}{arXiv:2207.08885 [hep-th]}.

\bibitem{ADM}
R. L. Arnowitt, S. Deser and C. W. Misner,
{\it The Dynamics of general relativity},
\href{https://doi.org/10.1007/s10714-008-0661-1}{Gen. Rel. Grav. {\bf 40}, 1997 (2008)}.

\bibitem{Baierlein}
R.~F.~Baierlein, D.~H.~Sharp and J.~A.~Wheeler,
{\it Three-Dimensional Geometry as Carrier of Information about Time},
\href{https://doi.org/10.1103/PhysRev.126.1864}{Phys. Rev. \textbf{126}, 1864 (1962)}.

\bibitem{strings1}
Y. Nambu,
Lectures on the Copenhagen Summer Symposium (1970), unpublished.

\bibitem{strings2}
T. Goto,
{\it Relativistic Quantum Mechanics of One-Dimensional Mechanical Continuum and Subsidiary Condition of Dual Resonance Model},
\href{https://doi.org/10.1143/PTP.46.1560}{Progr. Theor. Phys. {\bf 46}, 1560 (1971)}.

\bibitem{strings3}
A.M. Polyakov,
{\it Quantum geometry of bosonic strings},
\href{https://doi.org/10.1016/0370-2693(81)90743-7}{Phys. Lett. B {\bf 103}, 207 (1981)}.

\bibitem{strings4}
J.~Polchinski,
{\it String theory Vol. 1: An introduction to the bosonic string},
Cambridge University Press, Cambridge, England (2007).

\bibitem{StueckTrick}
E. C. G. Stueckelberg,
{\it Die Wechselwirkungs Kraefte in der Elektrodynamik und in der Feldtheorie der Kernkraefte (I)},
Helv. Phys. Acta 11, 225 (1938); 
{\it Die Wechselwirkungs Kraefte in der Elektrodynamik und in der Feldtheorie der Kernkraefte (II)},
Helv. Phys. Acta 11, 299 (1938);
{\it Die Wechselwirkungs Kraefte in der Elektrodynamik und in der Feldtheorie der Kernkraefte (III)},
Helv. Phys. Acta 11, 312 (1938).

\bibitem{StueckTrick1}
H.~Ruegg and M.~Ruiz-Altaba,
{\it The Stueckelberg field},
\href{https://doi.org/10.1142/S0217751X04019755}{Int. J. Mod. Phys. A \textbf{19}, 3265 (2004)}.

\bibitem{StueckTrick2}
S.~L.~Lyakhovich,
{\it General method for including Stueckelberg fields},
\href{https://doi.org/10.1140/epjc/s10052-021-09256-9}{Eur. Phys. J. C \textbf{81}, 472 (2021)}.

\bibitem{SalisburyMomentum}
D.~Salisbury,
{\it Leon Rosenfeld and the challenge of the vanishing momentum in quantum electrodynamics},
\href{https://doi.org/10.1016/j.shpsb.2009.08.002}{Stud. Hist. Phil. Sci. B \textbf{40}, 363 (2009)}.

\bibitem{NambuEM}
Y.~Nambu,
{\it Quantum electrodynamics in nonlinear gauge},
\href{https://doi.org/10.1143/PTPS.E68.190}{Prog. Theor. Phys. Suppl. E \textbf{68}, 190 (1968)}.

\bibitem{Fermi}
E.~Fermi,
{\it Quantum Theory of Radiation},
\href{https://doi.org/10.1103/RevModPhys.4.87}{Rev. Mod. Phys. \textbf{4}, 87 (1932)}.

\bibitem{Proca}
A. Proca,
{\it Sur la th\'eorie ondulatoire des \'electrons positifs et n\'egatifs}, 
J. de Phys. et le Radium {\bf7}, 347 (1936).

\bibitem{MTW}
C.~W.~Misner, K.~S.~Thorne and J.~A.~Wheeler,
{\it Gravitation},
W. H. Freeman (1973).

\bibitem{DeWittTrilogy1}
B. S. DeWitt,
{\it Quantum Theory of Gravity. I. The Canonical Theory},
\href{https://doi.org/10.1103/PhysRev.160.1113}{Phys. Rev. {\bf 160}, 1113 (1967)}.

\bibitem{we}
A. O. Barvinsky and A. Yu. Kamenshchik,
{\it Darkness without dark matter and energy - generalized unimodular gravity},
\href{https://doi.org/10.1016/j.physletb.2017.09.045}{Phys. Lett. B {\bf 774}, 59 (2017)}.

\bibitem{Synge}
J. L. Synge,
{\it Relativity: the General Theory},
Amsterdam: North-Holland (1961).

\bibitem{Ellis}
G.~F.~R.~Ellis and D.~Garfinkle,
{\it The Synge G-Method: cosmology, wormholes, firewalls, geometry},
\href{https://doi.org/10.1088/1361-6382/ad2f14}{Class. Quant. Grav. \textbf{41}, 077002 (2024)}.

\bibitem{Barv}
A.~O.~Barvinsky, N.~Kolganov, A.~Kurov and D.~Nesterov,
{\it Dynamics of the generalized unimodular gravity theory},
\href{https://doi.org/10.1103/PhysRevD.100.023542}{Phys.\ Rev.\ D {\bf 100}, 023542 (2019)}.

\bibitem{Barv1}
A.~O.~Barvinsky and N.~Kolganov,
{\it Inflation in generalized unimodular gravity},
\href{https://doi.org/10.1103/PhysRevD.100.123510}{Phys.\ Rev.\ D {\bf 100}, 123510 (2019)}.
  
\bibitem{Kam-Tron-Ven}
A. Yu. Kamenshchik, A. Tronconi and G. Venturi,
{\it Generalized unimodular gravity in Friedmann and Kantowski\textendash{}Sachs universes}, 
\href{https://doi.org/10.1134/S0021364020080032}{JETP Lett. {\bf 111}, 416 (2020)}.

\bibitem{unfree}
D.~S.~Kaparulin and S.~L.~Lyakhovich,
{\it A note on unfree gauge symmetry},
\href{https://doi.org/10.1016/j.nuclphysb.2019.114735}{Nucl.\ Phys.\ B {\bf 947}, 114735 (2019)}.
  
\bibitem{unfree1}
D.~S.~Kaparulin and S.~L.~Lyakhovich,
{\it Unfree gauge symmetry in the BV formalism},
\href{https://doi.org/10.1140/epjc/s10052-019-7233-2}{Eur.\ Phys.\ J.\ C {\bf 79}, 718 (2019)}.

\bibitem{unfree2}
V.~A.~Abakumova, I.~Y.~Karataeva and S.~L.~Lyakhovich,
{\it Unfree gauge symmetry in the Hamiltonian formalism},
\href{https://doi.org/10.1016/j.physletb.2020.135208}{Phys.\ Lett.\ B {\bf 802}, 135208 (2020)}.

\bibitem{unfree3}
V.~A.~Abakumova  and S.~L.~Lyakhovich,
{\it Unfree gauge symmetry},
\href{https://doi.org/10.1134/S1063779623050179}{Phys. Part. Nucl. \textbf{54}, 950 (2023)}.

\bibitem{Barv-Nest}
A.~O.~Barvinsky and D.~V.~Nesterov,
{\it Restricted gauge theory formalism and unimodular gravity},
\href{https://doi.org/10.1103/PhysRevD.108.065004}{Phys. Rev. D \textbf{108}, 065004 (2023)}.

\bibitem{Tdiff}
D. Jaramillo-Garrido, A. L. Maroto and P. Mart\'in-Moruno,
{\it TDiff in the Dark: Gravity with a scalar field invariant under transverse diffeomorphisms},
\href{https://doi.org/10.1007/JHEP03(2024)084}{Journal of High Energy Physics \textbf{03}, 084 (2024)}.
 
\bibitem{Tdiff1} 
A.G. Bello-Morales, A.L. Maroto,
{\it Cosmology in gravity models with broken diffeomorphisms}, 
\href{https://doi.org/10.1103/PhysRevD.109.043506}{Phys. Rev. D \textbf{109}, 043506 (2024)}.

\bibitem{Percacci}
S. Gielen, R. de Leon Ardon and R. Percacci,
{\it Gravity with more or less gauging},
\href{https://doi.org/10.1088/1361-6382/aadbd1}{Class. Quantum Grav. {\bf 35}, 195009 (2018)}.

\bibitem{Redef}
A.~Y.~Kamenshchik, A.~Tronconi and G.~Venturi,
{\it Quantum Cosmology and the Evolution of Inflationary Spectra},
\href{https://doi.org/10.1103/PhysRevD.94.123524}{Phys. Rev. D \textbf{94}, 123524 (2016)}.

\bibitem{Burlan}
D. E. Burlankov,
{\it Physics of Space and Time} (in Russian), Nizhny Novgorod (2014).

\bibitem{Schutz}
B. F. Schutz,
{\it Perfect fluids in general relativity: Velocity potentials and a variational principle},
\href{https://doi.org/10.1103/PhysRevD.2.2762}{Phys. Rev. D {\bf2}, 2762 (1970)};
{\it Hamiltonian Theory of a Relativistic Perfect Fluid},
\href{https://doi.org/10.1103/PhysRevD.4.3559}{Phys. Rev. D \textbf{4}, 3559 (1971)}.

\bibitem{LapRuba}
V. G. Lapchinskii and V.A. Rubakov,
{\it Quantum gravitation: Quantization of the Friedmann model},
\href{https://doi.org/10.1007/BF01036991}{Theor. Math. Phys. {\bf33}, 1076 (1977)}.

\bibitem{Alvarenga}
F.~G.~Alvarenga, A.~B.~Batista, J.~C.~Fabris and S.~V.~B.~Goncalves,
{\it Troubles with quantum anistropic cosmological models: Loss of unitarity},
\href{https://doi.org/10.1023/A:1025735202959}{Gen. Rel. Grav. \textbf{35}, 1659 (2003)}.

\bibitem{Almeida}
C.~R.~Almeida, A.~B.~Batista, J.~C.~Fabris and P.~R.~L.~V.~Moniz,
{\it Quantum cosmology with scalar fields: self-adjointness and cosmological scenarios},
\href{https://doi.org/10.1134/S0202289315030020}{Grav. Cosmol. \textbf{21}, 191 (2015)}.

\bibitem{Molinari}
P.~A.~P.~Molinari, P.~C.~M.~Delgado, R.~F.~Pinheiro and N.~Pinto-Neto,
{\it Radiation-dominated bouncing model with slow contraction and inflation},
\href{https://doi.org/10.1103/PhysRevD.109.043531}{Phys. Rev. D \textbf{109}, 043531 (2024)}.

\bibitem{Marolf}
D.~Marolf,
{\it Group averaging and refined algebraic quantization: Where are we now?},
{The Ninth Marcel Grossmann Meeting (2000)},
\href{https://arxiv.org/abs/gr-qc/0011112}{arXiv:gr-qc/0011112 [gr-qc]}.

\bibitem{Giulini}
D.~Giulini and D.~Marolf,
{\it A Uniqueness theorem for constraint quantization},
\href{https://doi.org/10.1088/0264-9381/16/7/322}{Class. Quant. Grav. \textbf{16}, 2489 (1999)};
{\it On the generality of refined algebraic quantization},
\href{https://doi.org/10.1088/0264-9381/16/7/321}{Class. Quant. Grav. \textbf{16}, 2479 (1999)}.

\bibitem{Kara}
I.~Y.~Karataeva and S.~L.~Lyakhovich,
{\it Gauge symmetry of unimodular gravity in Hamiltonian formalism},
\href{https://doi.org/10.1103/PhysRevD.105.124006}{Phys. Rev. D \textbf{105}, 124006 (2022)}.

\bibitem{DelGrosso}
L.~Del Grosso, D.~E.~Kaplan, T.~Melia, V.~Poulin, S.~Rajendran and T.~L.~Smith,
{\it Cosmological Consequences of Unconstrained Gravity and Electromagnetism},
\href{https://arxiv.org/abs/2405.06374}{arXiv:2405.06374 [hep-ph]}.

\bibitem{MagueijoNovo}
J.~Magueijo,
{\it Dark matter and spacetime symmetry restoration},
\href{https://doi.org/10.1103/PhysRevD.109.124026}{Phys. Rev. D \textbf{109} 124026 (2024)}.

\bibitem{MagueijoNovo2}
J.~Magueijo,
{\it Space-time symmetry breaking on non-geodesic leaves and a new form of matter},
\href{https://arxiv.org/abs/2406.17428}{arXiv:2406.17428 [gr-qc]}.

\bibitem{Mukohyama}
S.~Mukohyama,
{\it Dark matter as integration constant in Horava-Lifshitz gravity},
\href{https://doi.org/10.1103/PhysRevD.80.064005}{Phys. Rev. D \textbf{80}, 064005 (2009)}.

\bibitem{Kobakhidze}
A.~Kobakhidze,
{\it On the infrared limit of Horava's gravity with the global Hamiltonian constraint},
\href{https://doi.org/10.1103/PhysRevD.82.064011}{Phys. Rev. D \textbf{82}, 064011 (2010)}.

\bibitem{Sundermeyer}
K.~Sundermeyer, {\it Constrained Dynamics With Applications To Yang-Mills Theory, General Relativity, Classical Spin, Dual String Model}, Lecture Notes in Physics Vol 169, edited by H. Araki, J. Ehlers, K. Hepp, R. Kippenhahn, H. A. Weidenm\"{u}ller and J. Zittartz (Springer, Berlin, 1982).

\end{thebibliography}
\end{document}